\documentclass[english]{IEEEtran}
\usepackage[T1]{fontenc}
\usepackage[latin9]{inputenc}
\usepackage{xcolor}
\usepackage{pdfcolmk}
\usepackage{verbatim}
\usepackage{calc}
\usepackage{amsthm}
\usepackage{amsmath}
\usepackage{amssymb}
\usepackage{graphicx}
\PassOptionsToPackage{normalem}{ulem}
\usepackage{ulem}

\makeatletter

\providecommand{\tabularnewline}{\\}
\providecolor{lyxadded}{rgb}{0,0,1}
\providecolor{lyxdeleted}{rgb}{1,0,0}

  \theoremstyle{definition}
  \newtheorem{defn}{\protect\definitionname}
  \theoremstyle{remark}
  \newtheorem{rem}{\protect\remarkname}
  \theoremstyle{definition}
  \newtheorem{example}{\protect\examplename}
  \theoremstyle{plain}
  \newtheorem{lem}{\protect\lemmaname}
  \theoremstyle{plain}
  \newtheorem{thm}{\protect\theoremname}

\usepackage{cite}
\usepackage{times}

\newtheorem{assumption}{Assumption}
\newtheorem{challenge}{Challenge}

\@ifundefined{showcaptionsetup}{}{%
 \PassOptionsToPackage{caption=false}{subfig}}
\usepackage{subfig}
\makeatother

\usepackage{babel}
\providecommand{\definitionname}{Definition}
\providecommand{\examplename}{Example}
\providecommand{\lemmaname}{Lemma}
\providecommand{\remarkname}{Remark}
\providecommand{\theoremname}{Theorem}

\begin{document}

\title{Hierarchical Interference Mitigation for Massive MIMO Cellular Networks}

\author{{\normalsize{}An Liu, }\textit{\normalsize{}Member IEEE}{\normalsize{},
and Vincent Lau,}\textit{\normalsize{} Fellow IEEE}{\normalsize{},\\Department
of Electronic and Computer Engineering, Hong Kong University of Science
and Technology}}
\maketitle
\begin{abstract}
We propose a hierarchical interference mitigation scheme for massive
MIMO cellular networks. The MIMO precoder at each base station (BS)
is partitioned into an \textit{inner precoder} and an \textit{outer
precoder}. The inner precoder controls the intra-cell interference
and is adaptive to local channel state information (CSI) at each BS
(CSIT). The outer precoder controls the inter-cell interference and
is adaptive to channel statistics. Such hierarchical precoding structure
reduces the number of pilot symbols required for CSI estimation in
massive MIMO downlink and is robust to the backhaul latency. We study
joint optimization of the outer precoders, the user selection, and
the power allocation to maximize a general concave utility which has
no closed-form expression. We first apply random matrix theory to
obtain an approximated problem with closed-form objective. We show
that the solution of the approximated problem is asymptotically optimal
with respect to the original problem as the number of antennas per
BS grows large. Then using the hidden convexity of the problem, we
propose an iterative algorithm to find the optimal solution for the
approximated problem. We also obtain a low complexity algorithm with
provable convergence. Simulations show that the proposed design has
significant gain over various state-of-the-art baselines.\end{abstract}

\section{Introduction}

Massive MIMO is regarded as a promising technology in future wireless
networks due to its high spectrum and energy efficiency \cite{Marzetta_SPM12_LargeMIMO}.
The large spatial degree of freedom (DoF) of massive MIMO systems
can contribute to (i) spatial multiplexing gains for intra-cell users
per BS (MU-MIMO) as well as (ii) inter-cell interference mitigation
between the BSs via linear precoders at the BSs. In \cite{Peel_TOC05_RCI},
zero-forcing (ZF) and regularized zero-forcing (RZF) have been proposed
for spatial multiplexing of data streams to intra-cell users. More
complicated linear precoding schemes based on duality \cite{Boche_05TSP_MISOSINR}
or semidefinite relaxing (SDR) \cite{Gershman_SPM2010_SDRBF} have
also been proposed to achieve a better performance. On the other hand,
the inter-cell interference mitigation between BSs is more complicated.
One commonly adopted approach to mitigate the inter-cell interference
is the coordinated MIMO strategy \cite{Foschini_POC06_CordMIMO},
which performs joint precoding among the BSs using the global real-time
CSIs shared among the BSs. Alternatively, cooperative MIMO techniques
can also be exploit to mitigate inter-cell interference by sharing
both real-time CSI and payload data among the concerned BSs \cite{zhang2004cochannel}. 

However, these conventional spatial multiplexing and interference
mitigation techniques cannot be applied directly to massive MIMO cellular
networks due to the following reasons. First, the MU-MIMO precoding
requires real-time local CSIT at the BS. However, the amount of pilot
symbols for channel estimation is limited by the coherence time of
the channel and it is practically infeasible to obtain good CSI quality
when each BS is equipped with a massive MIMO array. Second, the existing
inter-cell interference mitigation methods such as cooperative and
coordinated MIMO require real-time global CSIT, which is difficult
to achieve in practice due to the backhaul latency%
\footnote{For example, the X2 interface in LTE systems has a typical latency
of 10ms or more between BSs.%
}. Hence, the performance of these schemes is very sensitive to CSIT
errors due to outdatedness.

In this paper, we address the above issues by proposing a hierarchical
interference mitigation scheme for massive MIMO cellular networks.
In the proposed scheme, the MIMO precoder at each BS is partitioned
into an \textit{inner precoder} and an \textit{outer precoder} as
illustrated in Fig. \ref{fig:HybridPC}. The inner precoder is used
to support MU-MIMO (control intra-cell interference and capture the
spatial multiplexing gain) at each BS and it is adaptive to real-time
local CSIT. The outer precoder can leverage on the remaining spatial
DoF to mitigate the inter-cell interference by restricting the transmitted
signal at each BS into a subspace and is adaptive to long-term channel
statistics%
\footnote{Due to local scattering effects \cite{Tse_05book_wireless_communication},
the MIMO spatial channels are not isotropic and precoding based on
statistical information can be quite effective to control / mitigate
the inter-cell interference as illustrated in Example \ref{exm:Txspace-motivation}.%
}. Such hierarchical precoding structure simultaneously resolves both
the aforementioned practical challenges. For instance, the issue of
insufficient pilot symbols for real-time local CSI estimation is resolved
because the BS only needs to estimate the CSI within the subspace
determined by the outer precoder, which is of a much smaller dimension
than the number of antennas. Furthermore, the outer precoder is adaptive
to the long-term channel statistics, which is insensitive to backhaul
latency. As a result, the proposed \textit{hierarchical precoding}
framework exploits the spatial DoF to simultaneously achieve spatial
multiplexing per BS and inter-cell interference mitigation without
expensive backhaul signaling requirement. We consider joint optimization
of the outer precoders, the user selection, and the power allocation
to maximize a general concave utility function of the average data
rates of users. The following first-order challenges need to be addressed.
\begin{itemize}
\item \textbf{Lack of Closed-Form Optimization Objective}: The average data
rate of each user involves stochastic expectation over CSI realizations
and it does not have closed form characterization.
\item \textbf{Complex Coupling between User Selection and Outer Precoding}:
The outer precoder will affect the admissible user set%
\footnote{For example, a user cannot be scheduled if its channel vector does
not lie in the subspace spanned by the outer precoder.%
}. On the other hand, the optimization of outer precoder also depends
on user selection because the outer precoder only needs to suppress
the interference to the selected users in other BSs. 
\item \textbf{Combinatorial Optimization Problem}: The user selection problem
with hierarchical precoding in the massive MIMO cellular networks
is combinatorial with exponential complexity w.r.t. the total number
of users.
\end{itemize}

To address the above challenges, we first apply the random matrix
theory to obtain an approximated problem with closed-form objective.
Then using the hidden convexity of the problem, we propose an iterative
algorithm to find the optimal solution for the approximated problem.
We also obtain a low complexity algorithm with provable convergence.
Finally, we illustrate with simulation that the proposed design achieves
significant performance gain compared with various state-of-the-art
baselines under various signaling backhaul latency.

\textit{Notation}\emph{s}: The superscripts $\left(\cdot\right)^{T}$
and $\left(\cdot\right)^{\dagger}$ denote transpose and Hermitian
respectively. For a set $\mathcal{S}$, $\left|\mathcal{S}\right|$
denotes the cardinality of $\mathcal{S}$. The operator $\textrm{diag}\left(\mathbf{a}\right)$
represents a diagonal matrix whose diagonal elements are the elements
of vector $\mathbf{a}$. The notation $\mathbb{U}^{M\times N}$ denote
the set of all $M\times N$ semi-unitary matrices. Let $1\left(\cdot\right)$
denote the indication function such that $1\left(E\right)=1$ if the
event $E$ is true and $1\left(E\right)=0$ otherwise. $\textrm{span}\left(\mathbf{A}\right)$
represents the subspace spanned by the columns of a matrix $\mathbf{A}$
and $\textrm{orth}\left(\mathbf{A}\right)$ represents a set of orthogonal
basis of $\textrm{span}\left(\mathbf{A}\right)$. $\left\Vert \mathbf{A}\right\Vert $
is the spectral radius of $\mathbf{A}$.

\section{System Model\label{sec:System-Model}}

\subsection{Massive MIMO Cellular Network}

\begin{figure}
\begin{centering}
\subfloat[A massive MIMO cellular network with $2$ BSs and $5$ users.]{\centering{}\includegraphics[width=60mm]{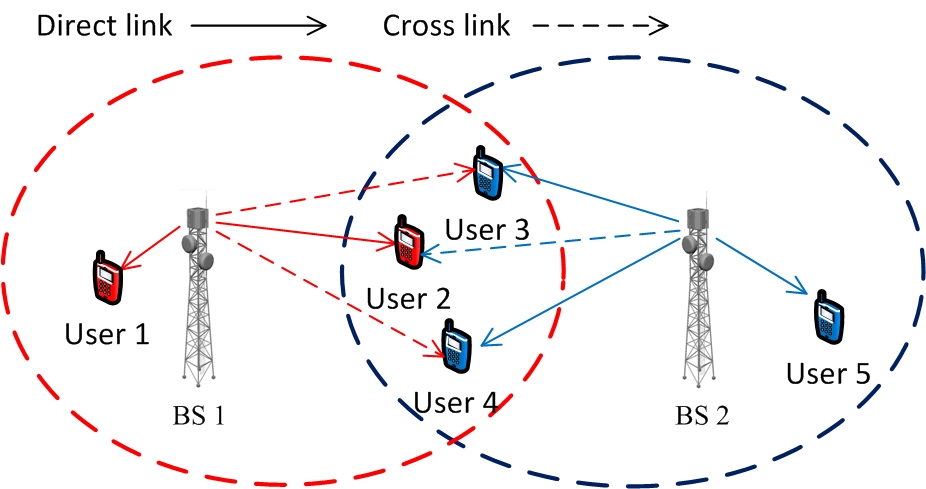}}
\par\end{centering}

\begin{centering}
\subfloat[The corresponding topology graph $\mathcal{G}_{T}\left(\mathbf{\Theta}\right)=\left\{ \mathcal{B},\mathcal{U},\mathcal{E}\right\} $,
where $\mathcal{B}=\left\{ 1,2\right\} $, $\mathcal{U}=\left\{ 1,2,3,4,5\right\} $
and $\mathcal{E}=\left\{ \left(1,1\right),\left(1,2\right),\left(1,3\right),\left(1,4\right),\left(2,2\right),\left(2,3\right),\left(2,4\right),\left(2,5\right)\right\} $.]{\centering{}\includegraphics[width=60mm]{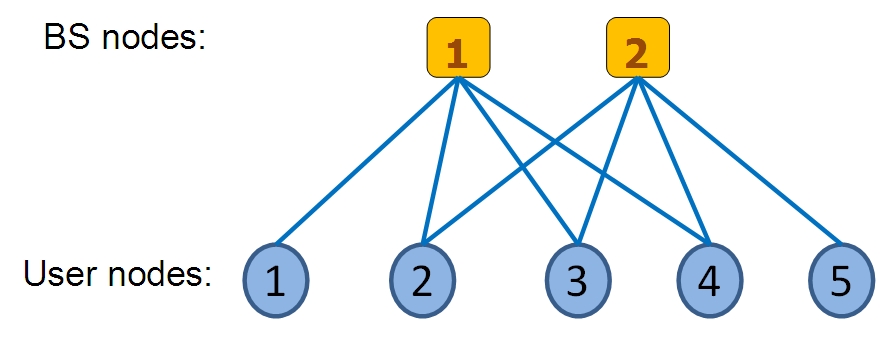}}
\par\end{centering}

\protect\caption{\label{fig:sysmodel}{\small{}An example of massive MIMO cellular
network and the corresponding topology graph.}}
\end{figure}
Consider the downlink of a massive MIMO cellular network with $N$
BSs and $K$ single-antenna users as illustrated in Fig. \ref{fig:sysmodel}
for $N=2$ and $K=5$. Each BS has $M$ antennas with $M$ much larger
than the number of the associated users. Denote $\mathbf{h}_{k,n}\in\mathbb{C}^{M}$
as the channel vector between BS $n$ and user $k$. The channel fading
process is modeled as $\mathbf{h}_{k,n}\left(t\right)=\sqrt{M}\mathbf{\Theta}_{k,n}^{1/2}\left(t\right)\mathbf{z}_{k,n}\left(t\right),\:\forall k,n$,
where $\mathbf{z}_{k,n}\left(t\right)\in\mathbb{C}^{M}$ has i.i.d.
complex entries of zero mean and variance $1/M$; and $\mathbf{\Theta}_{k,n}\left(t\right)\in\mathbb{C}^{M\times M}$
is the spatial correlation matrix between BS $n$ and user $k$. The
random process $\mathbf{z}_{k,n}\left(t\right)$ is quasi-static within
a time slot but i.i.d. w.r.t. time slots, user and BS indices $\left(t,k,n\right)$.
The spatial correlation process $\mathbf{\Theta}_{k,n}\left(t\right)$
is assumed to be a slow ergodic process (i.e., $\mathbf{\Theta}_{k,n}\left(t\right)$
remains constant for a large number of time slots) according to a
general distribution. As such, the CSI is divided into instantaneous
CSI $\mathbf{H}=\left\{ \mathbf{h}_{k,n}\right\} $ and global statistical
information $\mathbf{\Theta}\triangleq\left\{ \mathbf{\Theta}_{k,n}\right\} $
(spatial correlation matrices). Due to local scattering \cite{Tse_05book_wireless_communication},
the spatial correlation matrices of different users in cell $n$ is
usually different. However, if the coverage area of a BS is partitioned
into $N_{c}$ small sub-areas, it is reasonable to assume that any
two users collocated in the same sub-area have almost the same spatial
correlation matrices. This motivates us to consider the following
locally-clustered spatial channel model.

\begin{assumption}[Locally-clustered Spatial Channel]\label{asm:Chmodel}The
spatial correlation matrices $\left\{ \mathbf{\Theta}_{k,n},\forall k\right\} $
associated with BS $n$ belongs to a finite set $\mathbf{\Psi}_{n}$
with the size $\left|\mathbf{\Psi}_{n}\right|=N_{c}$. Furthermore,
due to the local spatial scattering \cite{Tse_05book_wireless_communication},
we have $\textrm{Rank}\left(\mathbf{\Theta}_{k,n}\right)<M,\:\forall k,n$.\hfill \IEEEQED

\end{assumption}

Assumption \ref{asm:Chmodel} is realistic because in practice, there
are only limited number of significant eigenvalues in a MIMO channel
(especially for large $M$). The massive MIMO cellular network can
be represented by a topology graph as define below.
\begin{defn}
[Network Topology Graph] For given spatial correlation matrices
$\mathbf{\Theta}$, define the \textit{topology graph} of the massive
MIMO cellular network as a bipartite graph $\mathcal{G}_{T}\left(\mathbf{\Theta}\right)=\left\{ \mathcal{B},\mathcal{U},\mathcal{E}\right\} $,
where $\mathcal{B}$ denotes the set of all BS nodes, $\mathcal{U}$
denotes the set of all user nodes, and $\mathcal{E}$ is the set of
all edges between the BSs and users. For each BS node $n$, let $\mathcal{U}_{n}$
denote the set of associated users and $\overline{\mathcal{U}}_{n}=\left\{ k:\: k\notin\mathcal{U}_{n},\left(k,n\right)\in\mathcal{E}\right\} $
denote the set of neighbor users. For each user node $k$, let $b_{k}$
denote the index of its serving BS and $\mathcal{B}_{k}=\left\{ n:\: n\neq b_{k},\left(k,n\right)\in\mathcal{E}\right\} $
denote the set of neighbor BSs.\hfill \IEEEQED
\end{defn}

Define $\textrm{E}\left[\left\Vert \mathbf{h}_{k,n}\right\Vert ^{2}\right]=\textrm{Tr}\left(\mathbf{\Theta}_{k,n}\right)$
as the \textit{path gain} between BS $n$ and user $k$. An edge between
a user node and a BS node in the network topology graph indicates
there is strong path gain between these two nodes. This is stated
formally below.
\begin{defn}
[Edge Set]\label{def:edgeset}For given $\mathbf{\Theta}$, there
is an edge $\left(k,n\right)\in\mathcal{E}$ between BS node $n\in\mathcal{B}$
and user node $k\in\mathcal{U}$ in the network topology graph $\mathcal{G}_{T}\left(\mathbf{\Theta}\right)=\left\{ \mathcal{B},\mathcal{U},\mathcal{E}\right\} $
if $\textrm{Tr}\left(\mathbf{\Theta}_{k,b_{k}}\right)<\theta\textrm{Tr}\left(\mathbf{\Theta}_{k,n}\right)$,
for some threshold $\theta>1$.\hfill \IEEEQED\end{defn}
\begin{rem}
In practical wireless networks, the data rate of each user is limited
by the available modulation and coding schemes (MCS) (e.g., the highest
MCS in LTE is 64QAM, no coding \cite{3gpp_Rel10_Phy}). If the path
gain between a user and a BS is sufficiently small compared to the
direct link path gain ($\theta$ times smaller than the direct link
path gain), the interference from this BS will have negligible effect
on the data rate of this user. Simulations show that the performance
of the proposed scheme is not sensitive to the choice of $\theta$
for a wide range of $\theta$ from $5$dB to $20$dB.
\end{rem}

An example of massive MIMO cellular network and the corresponding
topology graph is illustrated in Fig. \ref{fig:sysmodel}. For BS
$1$, the set of associated users is $\mathcal{U}_{1}=\left\{ 1,2\right\} $,
and the set of neighbor users is $\overline{\mathcal{U}}_{1}=\left\{ 3,4\right\} $.
For user $2$, the index of the serving BS is $b_{2}=1$ and the set
of neighbor BSs is $\mathcal{B}_{2}=\left\{ 2\right\} $. For user
$3$, the index of the serving BS is $b_{3}=2$ and the set of neighbor
BSs is $\mathcal{B}_{3}=\left\{ 1\right\} $.

At each time slot, linear precoding is employed at BS $n$ to support
simultaneous downlink transmissions to a set of scheduled users denoted
by $\mathcal{S}_{n}$. Let $\mathcal{S}=\cup_{n=1}^{N}\mathcal{S}_{n}$
denote the set of all the selected users and $\overline{\mathcal{S}}_{n}=\overline{\mathcal{U}}_{n}\cap\mathcal{S}$
denote the set of selected users who are neighbors of BS $n$. Note
that a user $k\in\overline{\mathcal{S}}_{n}$ can be potentially interfered
by BS $n$ because there is a cross link (edge) between BS $n$ and
a user $k\in\overline{\mathcal{S}}_{n}$. For example, consider the
massive MIMO cellular network in Fig. \ref{fig:sysmodel}. Suppose
the sets of selected users at the BSs are $\mathcal{S}_{1}=\left\{ 1,2\right\} $
and $\mathcal{S}_{2}=\left\{ 3,5\right\} $. Then, we have $\mathcal{S}=\mathcal{S}_{1}\cup\mathcal{S}_{2}=\left\{ 1,2,3,5\right\} $
and $\overline{\mathcal{S}}_{1}=\overline{\mathcal{U}}_{1}\cap\mathcal{S}=\left\{ 3\right\} $,
where $\overline{\mathcal{U}}_{1}=\left\{ 3,4\right\} $. Since user
$3$ has a cross link with BS $1$ as illustrated in Fig. \ref{fig:sysmodel}-(a),
it can be potentially interfered by BS $1$. Using the above notations,
the received signal for a user $k$ can be expressed as:
\begin{eqnarray*}
y_{k} & = & \mathbf{h}_{k,b_{k}}^{\dagger}\sqrt{p_{k}}\mathbf{v}_{k}s_{k}+\underset{\textrm{intracell\:\ interference}}{\underbrace{\mathbf{h}_{k,b_{k}}^{\dagger}\sum_{l\in\mathcal{S}_{b_{k}}\backslash\{k\}}\sqrt{p_{l}}\mathbf{v}_{l}s_{l}}}\\
 &  & +\underset{\textrm{intercell\:\ interference}}{\underbrace{\sum_{n\in\mathcal{B}_{k}}\mathbf{h}_{k,n}^{\dagger}\mathbf{V}_{n}\mathbf{P}_{n}\mathbf{s}_{n}}}+n_{k},
\end{eqnarray*}
where $s_{k}\sim\mathcal{CN}\left(0,1\right)$ is the data symbol,
$p_{k}$ is the power allocation and $\mathbf{v}_{k}$ is the precoding
vector of user $k$; $\mathcal{S}_{b_{k}}$ is the set of selected
users at BS $b_{k}$; $\mathbf{s}_{n}=\left[s_{l}\right]_{l\in\mathcal{S}_{n}}\in\mathbb{C}^{\left|\mathcal{S}_{n}\right|}$
is the data symbol vector at BS $n$; $\mathbf{P}_{n}=\textrm{diag}\left(\mathbf{p}_{n}\right)$
and $\mathbf{p}_{n}=\left[p_{l}\right]_{l\in\mathcal{S}_{n}}\in\mathbb{R}_{+}^{\left|\mathcal{S}_{n}\right|}$
is the power allocation vector at BS $n$; $\mathbf{V}_{n}=\left[\mathbf{v}_{l}\right]_{l\in\mathcal{S}_{n}}\in\mathbb{C}^{M\times\left|\mathcal{S}_{n}\right|}$
is the precoding matrix at BS $n$; and $n_{k}\sim\mathcal{CN}\left(0,1\right)$
is the AWGN noise.

\subsection{Hierarchical Interference Mitigation }

Conventional interference mitigation techniques for small scale MIMO
cellular networks such as MU-MIMO precoding, coordinated MIMO \cite{Foschini_POC06_CordMIMO},
or cooperative MIMO \cite{zhang2004cochannel}, cannot be applied
directly to massive MIMO cellular networks due to two practical challenges,
namely, the insufficient pilot symbols for CSI estimation and the
backhaul latency. To resolve these practical challenges, we propose
a novel hierarchical interference mitigation control, which can fully
utilize the large number of antennas to simultaneously mitigate the
inter-cell interference as well as realize the spatial multiplexing
gain per BS. Specifically, the interference mitigation strategy is
partitioned into \textit{long-term} and \textit{short-term} controls.
The short-term control is responsible to capture the spatial multiplexing
gain among the intra-cell users at each BS based on the local CSIT
only. On the other hand, the long-term control is responsible to mitigate
the inter-cell interference based on the global statistical information.
They are elaborated as follows.

\subsubsection{Hierarchical Precoding for Intra-cell and Inter-cell Interference
Mitigation}

We first use a simple example to illustrate the idea of hierarchical
precoding.
\begin{figure}
\begin{centering}
\includegraphics[width=70mm]{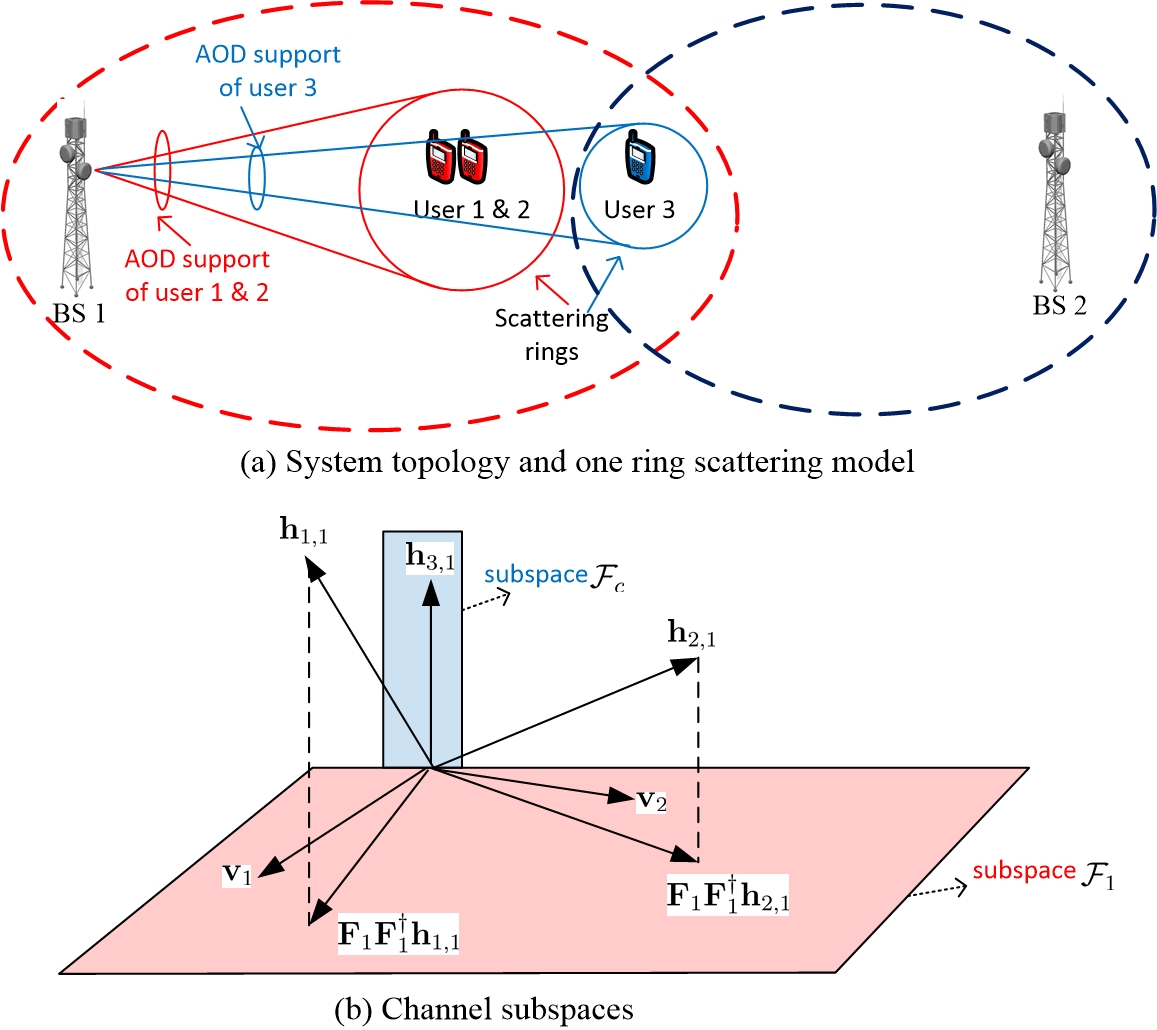}
\par\end{centering}

\protect\caption{\label{fig:Exmphych}{\small{}An example massive MIMO cellular network
with one ring scattering model to illustrate the role of spatial correlation
matrices $\mathbf{\Theta}$. The direct link channel subspace $\mathcal{F}$
is illustrated in Subfigure (b). It consists of two orthogonal subspaces
$\mathcal{F}_{c}$ and $\mathcal{F}_{1}$.}}
\end{figure}

\begin{example}
\label{exm:Txspace-motivation}Consider the massive MIMO cellular
network in Fig. \ref{fig:Exmphych}-(a). Each BS has $32$ antennas.
Consider the one ring model in \cite{Caire_TIT13_JSDM} for transmit
antenna correlation, where a user is surrounded by a ring of scatterers
such that the support of its Angle-of-Departure (AOD) distribution
are restricted to a certain region as illustrated in Fig. \ref{fig:Exmphych}-(a).
Assume that user 1 and user 2 share the same scattering ring and $\mathbf{h}_{1,1},\mathbf{h}_{2,1}$
are restricted in the same 4-dimensional subspace $\mathcal{F}$ (i.e.,
$\textrm{span}\left(\mathbf{\Theta}_{1,1}\right)=\textrm{span}\left(\mathbf{\Theta}_{2,1}\right)=\mathcal{F}$)
as illustrated in Fig. \ref{fig:Exmphych}-(b). Moreover, due to the
local scattering configuration as illustrated in Fig. \ref{fig:Exmphych}-(a),
the cross link channel vector $\mathbf{h}_{3,1}$ between BS 1 and
user 3 is restricted in a 2-dimensional subspace $\mathcal{F}_{c}\subset\mathcal{F}$
(i.e., $\textrm{span}\left(\mathbf{\Theta}_{3,1}\right)=\mathcal{F}_{c}\subset\mathcal{F}$).
We consider a hierarchical precoder structure for BS 1: $\mathbf{V}_{1}=\mathbf{F}_{1}\mathbf{G}_{1}$,
where $\mathbf{F}_{1}\in\mathbb{U}^{32\times2}$ with $\textrm{span}\left(\mathbf{F}_{1}\right)\subseteq\mathcal{F}$
is the outer precoder adaptive to the spatial correlation matrices
$\mathbf{\Theta}$ only, and $\mathbf{G}_{1}\in\mathbb{C}^{2\times2}$
is the inner precoder adaptive to the local CSI (the effective channel
$\widetilde{\mathbf{H}}_{1}=\left[\mathbf{F}_{1}^{\dagger}\mathbf{h}_{1,1},\mathbf{F}_{1}^{\dagger}\mathbf{h}_{2,1}\right]^{\dagger}\in\mathbb{C}^{2\times2}$).
\end{example}

We have the following observations from Example \ref{exm:Txspace-motivation}.

\textbf{Role of spatial correlation matrices $\mathbf{\Theta}$:}
The knowledge of the spatial correlation matrices \textbf{$\mathbf{\Theta}$
}can be exploited to design the\textit{ }outer precoder $\mathbf{F}_{1}$
to eliminate the inter-cell interference to user $3$. Specifically,
this can be achieved by setting $\mathbf{F}_{1}$ to be a set of orthogonal
basis of a 2-dimensional subspace $\mathcal{F}_{1}$, where $\mathcal{F}_{1}\subset\mathcal{F}$
and $\mathcal{F}_{1}\perp\mathcal{F}_{c}$ as illustrated in Fig.
\ref{fig:Exmphych}-(b).

\textbf{Role of local CSI $\widetilde{\mathbf{H}}_{1}$:} The knowledge
of local real-time instantaneous CSI \textbf{$\widetilde{\mathbf{H}}_{1}$}
can be exploited to design the\textit{ }inner precoder $\mathbf{G}_{1}$
to realize the spatial multiplexing gain at BS 1.%

\begin{figure}
\begin{centering}
\includegraphics[width=70mm]{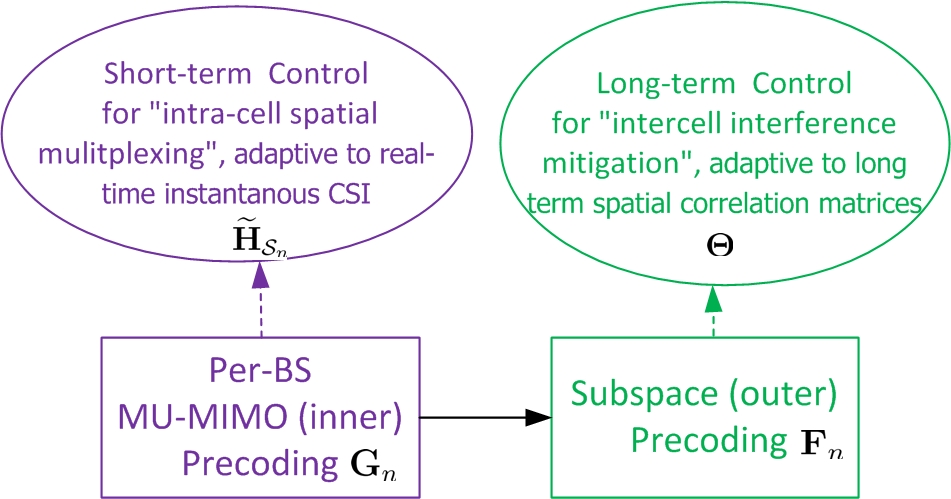}
\par\end{centering}

\protect\caption{\label{fig:HybridPC}{\small{}An illustration of hierarchical precoder
structure.}}
\end{figure}

For general massive MIMO cellular networks, we propose a hierarchical
precoder $\mathbf{V}_{n}=\mathbf{F}_{n}\mathbf{G}_{n}$ for each BS
$n$ as illustrated in Fig. \ref{fig:HybridPC}. The\textit{ outer
precoder} $\mathbf{F}_{n}\in\mathbb{U}^{M\times M_{n}}$ with $M_{n}<M$
(we let $\mathbf{F}_{n}=\mathbf{0}$ if $M_{n}=0$) is used to eliminate
the inter-cell interference and is adaptive to the global statistical
information $\mathbf{\Theta}$. The inner precoder $\mathbf{G}_{n}\in\mathbb{C}^{M_{n}\times\left|\mathcal{S}_{n}\right|}$
is used to realize the spatial multiplexing gain at each BS and is
adaptive to the local real-time CSI $\widetilde{\mathbf{H}}_{\mathcal{S}_{n}}\triangleq\left[\mathbf{F}_{n}^{\dagger}\mathbf{h}_{k,n}\right]_{k\in\mathcal{S}_{n}}^{\dagger}\in\mathbb{C}^{\left|\mathcal{S}_{n}\right|\times M_{n}}$.
Define $\mathbf{F}=\left\{ \mathbf{F}_{1},...,\mathbf{F}_{N}\right\} $
as the set of outer precoders for all BSs. By properly choosing $\mathbf{F}$
(equation (\ref{eq:ICIZF})), one can eliminate the inter-cell interference
as illustrated in Example \ref{exm:Txspace-motivation}.
\begin{rem}
Physically, the rank $M_{n}$ of $\mathbf{F}_{n}$ means the number
of data streams for spatial multiplexing at BS $n$. Due to limited
spatial scattering \cite{Tse_05book_wireless_communication}, a BS
with say $M=100$ antennas does not mean it can support spatial multiplexing
of 100 data streams. In practice, there are just a few (say $10$)
significant eigenchannels despite having 100 antennas, and having
$M_{n}=10$ spatially multiplexed data streams already capture most
of the spatial multiplexing advantage. The remaining spatial DoFs
can be used for inter-cell interference mitigation. 
\end{rem}

For a given outer precoder $\mathbf{F}$, we consider regularized
zero-forcing (RZF) inner precoder with a parameter $\nu$. The RZF
precoder is easy to implement and is asymptotically optimal for $M,\left|\mathcal{S}_{n}\right|\rightarrow\infty$
\cite{Hanly_IT11s_Multicell_RMT}. Moreover, we can apply the technique
of deterministic equivalent (DE) for RZF precoding in \cite{Wagner_TIT12s_LargeMIMO}
to facilitate the algorithm design. For convenience, define the composite
channel from BS $n$ to any subset of users $\mathcal{U}^{S}\subseteq\mathcal{U}$
as $\mathbf{H}_{\mathcal{U}^{S}}=\left[\mathbf{h}_{l,n}\right]_{l\in\mathcal{U}^{S}}^{\dagger}\in\mathbb{C}^{\left|\mathcal{U}^{S}\right|\times M}$.
If the inter-cell interference is completely eliminated by the outer
precoders $\mathbf{F}$, the RZF inner precoder is given by 
\begin{eqnarray}
\mathbf{G}_{n} & = & \left(\mathbf{F}_{n}^{\dagger}\mathbf{H}_{\mathcal{S}_{n}}^{\dagger}\mathbf{H}_{\mathcal{S}_{n}}\mathbf{F}_{n}+M\nu\mathbf{I}_{M_{n}}\right)^{-1}\mathbf{F}_{n}^{\dagger}\mathbf{H}_{\mathcal{S}_{n}}^{\dagger},\label{eq:RZFVn}
\end{eqnarray}
where $\nu>0$ is a fixed parameter for RZF. Note that $\nu$ is scaled
by $M$ to ensure that the matrix $\mathbf{F}_{n}^{\dagger}\mathbf{H}_{\mathcal{S}_{n}}^{\dagger}\mathbf{H}_{\mathcal{S}_{n}}\mathbf{F}_{n}+M\nu\mathbf{I}_{M_{n}}$
is well conditioned as $M,\left|\mathcal{S}_{n}\right|\rightarrow\infty$.

\subsubsection{Statistical User Scheduling and Power Allocation}

In \cite{Caire_ISIT09_USZF}, it has been observed that as the number
of antennas $M$ grows large, the role of multi-user diversity gain
(by selecting users based on instantaneous CSIT) becomes less and
less effective because of \textquotedblleft channel hardening\textquotedblright .
Moreover, the benefits of short timescale power allocation (i.e.,
the power allocation is adaptive to instantaneous CSIT) becomes asymptotically
negligible as $K,M\rightarrow\infty$ because the data rate of each
user converges almost surely to a deterministic function of the power
allocation vector $\mathbf{p}=\left[\mathbf{p}_{1}^{T},...,\mathbf{p}_{N}^{T}\right]^{T}$
as will be shown in Lemma \ref{lem:AsyWS}. As such, the user selection
set and power allocation is assumed to be adaptive to the global statistical
information $\mathbf{\Theta}$ only. Specifically, the user selection
and the outer precoder are chosen to satisfy the \textit{zero inter-cell
interference constraint}:
\begin{equation}
\mathbf{F}_{n}^{\dagger}\sum_{k\in\overline{\mathcal{S}}_{n}}\mathbf{\Theta}_{k,n}=\mathbf{0},\:\forall n,\label{eq:ICIZF}
\end{equation}
and the power allocation has to satisfy a per-BS power constraint
that will be elaborated later. Note that\textbf{ }the inter-cell interference
from BS $n$ to a user $k\in\overline{\mathcal{S}}_{n}$ can be expressed
as $M\mathbf{z}_{k,n}^{\dagger}\mathbf{\Theta}_{k,n}\mathbf{F}_{n}\mathbf{G}_{n}\mathbf{P}_{n}\mathbf{G}_{n}^{\dagger}\mathbf{F}_{n}^{\dagger}\mathbf{\Theta}_{k,n}\mathbf{z}_{k,n}$.
Since $\mathbf{\Theta}_{k,n}$ is positive semidefinite, setting\textbf{
$\mathbf{F}_{n}^{\dagger}\sum_{k\in\overline{\mathcal{S}}_{n}}\mathbf{\Theta}_{k,n}=\mathbf{0}$
}is equivalent to setting $\mathbf{F}_{n}^{\dagger}\mathbf{\Theta}_{k,n}=\mathbf{0},\:\forall k\in\overline{\mathcal{S}}_{n}$.
Hence, the constraint in (\ref{eq:ICIZF}) can be used to eliminate
the inter-cell interference to all users in the system. In (\ref{eq:ICIZF}),
we consider ZF criteria for inter-cell interference mitigation due
to its simplicity and asymptotic optimality at high SNR. Similar ZF
criteria has also been used in \cite{Caire_TIT13_JSDM} to design
pre-beamforming matrix based on spatial correlation matrices in single
cell systems.
\begin{rem}
[Implementation Considerations]\label{rem:LCSI}In the proposed
hierarchical interference mitigation, the long term controls such
as outer precoding, statistical user selection and power allocation
are implemented at a central node based on the global statistical
information about the channel ($\mathbf{\Theta}$), while the short
term control (inner precoding) is implemented locally at each BS based
on the local real-time instantaneous CSI knowledge ($\widetilde{\mathbf{H}}_{\mathcal{S}_{n}}$).
The proposed hierarchical precoding solution has several unique benefits
regarding implementation. (a) Robust to CSI signaling latency in backhaul,
(b) Resolve the issues of insufficient pilot and feedback overhead
for real-time CSIT estimation in massive MIMO systems. For instance,
the central node requires the global statistical information (spatial
correlation matrices) $\mathbf{\Theta}$ to compute the long term
controls $\left\{ \mathbf{F},\mathcal{S},\mathbf{p}\right\} $. The
spatial correlation matrices can be estimated via downlink training
using some standard covariance matrix estimation technique \cite{Mestre_TIT08_CovEst}
at the users and then fed back to the BSs. Since $\mathbf{\Theta}$
changes at a much slower time scale w.r.t. the time slot rate, such
a design requires substantially less signaling overhead compared to
the coordinated MIMO and is more robust w.r.t. to the backhaul latency.
On the other hand, BS $n$ only needs to know the local real-time
instantaneous CSI $\widetilde{\mathbf{H}}_{\mathcal{S}_{n}}\in\mathbb{C}^{\left|\mathcal{S}_{n}\right|\times M_{n}}$
for the inner precoder $\mathbf{G}_{n}$. This can be obtained via
downlink channel estimation and channel feedback using conventional
CSI signaling mechanisms in modern wireless systems such as LTE \cite{LTE}.
Since $M_{n}$ is substantially smaller than $M$, the issue of the
huge downlink pilot and CSI feedback signaling overhead in massive
MIMO is also alleviated by the hierarchical precoding design.
\end{rem}

\section{Optimization Formulation for Hierarchical Interference Mitigation\label{sec:Optimization-Formulation-for}}

We consider joint optimization of the outer precoders $\mathbf{F}$,
the user selection $\mathcal{S}$, and the power allocation $\mathbf{p}=\left[\mathbf{p}_{1}^{T},...,\mathbf{p}_{N}^{T}\right]^{T}$;
all of them are adaptive to the global statistical information $\mathbf{\Theta}$.
Define $\Gamma=\left\{ \mathbf{F},\mathcal{S},\mathbf{p}\right\} $
as a\textit{ composite control }variable. For given $\Gamma=\left\{ \mathbf{F},\mathcal{S},\mathbf{p}\right\} $
that satisfies (\ref{eq:ICIZF}), the instantaneous data rate (treating
interference as noise) of user $k$ is
\begin{equation}
r_{k}\left(\Gamma\right)=\textrm{log}\left(1+\frac{1\left(k\in\mathcal{S}\right)p_{k}\left|\mathbf{h}_{k,b_{k}}^{\dagger}\mathbf{v}_{k}\right|^{2}}{\sum_{l\in\mathcal{S}_{b_{k}}\backslash k}p_{l}\left|\mathbf{h}_{k,b_{k}}^{\dagger}\mathbf{v}_{l}\right|^{2}+1}\right),\label{eq:Irate}
\end{equation}
where the precoders $\left[\mathbf{v}_{l}\right]_{l\in\mathcal{S}_{n}}=\mathbf{F}_{n}\mathbf{G}_{n},\forall n$
with the inner precoder $\mathbf{G}_{n}$ given by (\ref{eq:RZFVn}).
The transmit power of BS $n$ is{\small{}
\begin{eqnarray}
P_{n}\left(\Gamma\right)=\;\;\;\;\;\;\;\;\;\;\;\;\;\;\;\;\;\;\;\;\;\;\;\;\;\;\;\;\;\;\;\;\;\;\;\;\;\;\;\;\;\;\;\;\;\;\;\;\;\;\;\;\;\;\;\;\;\;\;\;\;\;\;\;\;\;\;\;\;\;\;\nonumber \\
\textrm{Tr}\left(\mathbf{P}_{n}\mathbf{H}_{\mathcal{S}_{n}}\mathbf{F}_{n}\left(\mathbf{F}_{n}^{\dagger}\mathbf{H}_{\mathcal{S}_{n}}^{\dagger}\mathbf{H}_{\mathcal{S}_{n}}\mathbf{F}_{n}+M\nu\mathbf{I}_{M_{n}}\right)^{-2}\mathbf{F}_{n}^{\dagger}\mathbf{H}_{\mathcal{S}_{n}}^{\dagger}\right).
\end{eqnarray}
}Note that there may not always be enough spatial DoFs to eliminate
the inter-cell interference to all the users. Hence, for a fixed composite
control variable $\Gamma$, it is possible that only part of the users
can be scheduled for transmission. For fairness considerations, we
consider randomized control policy which realizes time-sharing between
several composite control variables as defined below.
\begin{defn}
[Randomized Control Policy]\label{def:Randomized-Control-PolicyA}A
randomized control policy $\Omega=\left\{ \Xi,\mathbf{q}\right\} $
consists of a set of composite control variables $\Xi\triangleq\left\{ \Gamma_{1},...,\Gamma_{\left|\Xi\right|}\right\} $
with $\left|\Xi\right|\leq K$ and a probability vector $\mathbf{q}\triangleq\left[q_{1},...,q_{\left|\Xi\right|}\right]^{T}$,
where the $j$-th composite control variable in $\Xi$ is $\Gamma_{j}=\left\{ \mathbf{F}\left(j\right),\mathcal{S}\left(j\right),\mathbf{p}\left(j\right)\right\} $;
and $\mathbf{q}$ satisfies $q_{j}\in\left[0,1\right],\forall j;\:\sum_{j=1}^{\left|\Xi\right|}q_{j}=1$.
At any time slot, the composite control variable $\Gamma_{j}$ is
used with probability $q_{j}$, i.e., the outer precoders, the user
selection set and the power allocation are respectively given by $\mathbf{F}\left(j\right)$,
$\mathcal{S}\left(j\right)$ and $\mathbf{p}\left(j\right)$ with
probability $q_{j}$. Moreover, define the set of feasible control
policies under per-BS power constraint $P_{c}$ as
\[
\Lambda\left(P_{c}\right)=\left\{ \Omega\triangleq\left\{ \Xi,\mathbf{q}\right\} :\:\Xi\subseteq\Xi^{\textrm{F}}\left(P_{c}\right)\right\} ,
\]
where $\Xi^{\textrm{F}}\left(P_{c}\right)=\left\{ \Gamma\in\Xi^{\textrm{A}}:\:\textrm{E}\left[\left.P_{n}\left(\Gamma\right)\right|\mathbf{\Theta}\right]\leq P_{c}\right\} $
is the set of feasible composite control variables, and $\Xi^{\textrm{A}}=\left\{ \Gamma:\:\mathbf{F}_{n}^{\dagger}\sum_{k\in\overline{\mathcal{S}}_{n}}\mathbf{\Theta}_{k,n}=\mathbf{0},\forall n\right.$
$\left.\:\textrm{and}\:\underset{k\in\mathcal{S}}{\textrm{sup}}p_{k}<\infty\right\} $
is the set of admissible composite control variables.\hfill \IEEEQED
\end{defn}

For given control policy $\Omega=\left\{ \Xi,\mathbf{q}\right\} $
and spatial correlation matrices $\mathbf{\Theta}$, the conditional
average data rate of user $k$ is:
\[
\overline{r}_{k}\left(\Omega|\mathbf{\Theta}\right)=\sum_{j=1}^{\left|\Xi\right|}q_{j}\textrm{E}\left[\left.r_{k}\left(\Gamma_{j}\right)\right|\mathbf{\Theta}\right].
\]

The network performance is characterized by a utility function $U\left(\overline{\mathbf{r}}\left(\Omega|\mathbf{\Theta}\right)\right)$,
where $\overline{\mathbf{r}}\left(\Omega|\mathbf{\Theta}\right)=\left[\overline{r}_{1}\left(\Omega|\mathbf{\Theta}\right),...,\overline{r}_{K}\left(\Omega|\mathbf{\Theta}\right)\right]^{T}$
is the conditional average rate vector. We make the following assumptions
on $U\left(\overline{\mathbf{r}}\right)$ ($\overline{\mathbf{r}}$
is a simplified notation for $\overline{\mathbf{r}}\left(\Omega|\mathbf{\Theta}\right)$).

\begin{assumption}[Assumptions on Utility]\label{asm:Ufun}The utility
function can be expressed as $U\left(\overline{\mathbf{r}}\right)\triangleq\sum_{k=1}^{K}w_{k}u\left(\overline{r}_{k}\right)$,
where $w_{k}\geq0$ is the weight for user $k$, $u\left(r\right)$
is assumed to be a twice differentiable, concave and increasing function
for all $r\geq0$. Moreover, $u\left(r\right)$ is L-Lipschitz, i.e.,
\[
\left|\frac{\partial u\left(r_{1}\right)}{\partial r}-\frac{\partial u\left(r_{2}\right)}{\partial r}\right|\leq L\left|r_{1}-r_{2}\right|,\forall r_{1},r_{2}\geq0,
\]
for some constant $L>0$.

\end{assumption}

The above utility function captures a lot of interesting cases:
\begin{itemize}
\item \textbf{Alpha-fair \cite{Mo_TACM00_alfafair}: }Alpha-fair can be
used to compromise between the fairness to users and the utilization
of resources. The utility function is%
\footnote{In the original alpha-fair utility function in \cite{Mo_TACM00_alfafair},
$\epsilon$ is equal to zero. In this paper, we set $\epsilon>0$
so that Assumption \ref{asm:Ufun} can be satisfied. Since $\epsilon$
is very small, it has negligible effect on the performance. The utility
function in (\ref{eq:alffair}) is also scaled by $\frac{1}{K}$ to
ensure that it is bounded as $K\rightarrow\infty$.%
}\textbf{ }
\begin{equation}
U\left(\overline{\mathbf{r}}\right)=\begin{cases}
\frac{1}{K}\sum_{k=1}^{K}\textrm{log}\left(\overline{r}_{k}+\epsilon\right), & \alpha=1,\\
\frac{1}{K}\sum_{k=1}^{K}\left(1-\alpha\right)^{-1}\left(\overline{r}_{k}+\epsilon\right)^{1-\alpha}, & \textrm{otherwise},
\end{cases}\label{eq:alffair}
\end{equation}
where $\epsilon>0$ is a small number. 
\item \textbf{Proportional Fair (PFS) \cite{Kelly_OPR98_PFS}:} This is
a special case of alpha-fair when $\alpha=1$. 
\end{itemize}

For a given topology graph $\mathcal{G}_{T}\left(\mathbf{\Theta}\right)=\left\{ \mathcal{B},\mathcal{U},\mathcal{E}\right\} $
and per-BS power constraint $P_{c}$, the problem of interference
mitigation via hierarchical precoding can be formulated as%
\footnote{Note that the set of feasible control policies $\Lambda\left(P_{c}\right)$
depends on $\mathcal{G}_{T}\left(\mathbf{\Theta}\right)$ since the
set of neighbor users $\overline{\mathcal{U}}_{n}$ of BS $n$ depends
on $\mathcal{G}_{T}\left(\mathbf{\Theta}\right)$.%
}:
\[
\mathcal{P}\left(\mathcal{G}_{T}\left(\mathbf{\Theta}\right)\right):\:\underset{\Omega}{\textrm{max}}\: U\left(\overline{\mathbf{r}}\left(\Omega|\mathbf{\Theta}\right)\right),\:\textrm{s.t.}\:\Omega\in\Lambda\left(P_{c}\right).
\]

Note that the conditional average rate $\overline{r}_{k}\left(\Omega|\mathbf{\Theta}\right)$
in the utility function and the conditional average power $\textrm{E}\left[\left.P_{n}\left(\Gamma\right)\right|\mathbf{\Theta}\right]$
in the constraint function of $\mathcal{P}\left(\mathcal{G}_{T}\left(\mathbf{\Theta}\right)\right)$
do not have closed form expressions. To make the problem tractable,
we need to address the following challenge.\vspace{-3bp}

\begin{flushleft}
\fbox{\begin{minipage}[t]{0.96\columnwidth}%
\begin{challenge}[Closed Form Approximation for $\mathcal{P}\left(\mathcal{G}_{T}\left(\mathbf{\Theta}\right)\right)$]Find
an approximated problem $\mathcal{P}_{E}\left(\mathcal{G}_{T}\left(\mathbf{\Theta}\right)\right)$
with closed form objective and constraints such that the solution
of $\mathcal{P}_{E}\left(\mathcal{G}_{T}\left(\mathbf{\Theta}\right)\right)$
is asymptotically optimal w.r.t. $\mathcal{P}\left(\mathcal{G}_{T}\left(\mathbf{\Theta}\right)\right)$
as $M$ grows large.\end{challenge}%
\end{minipage}}
\par\end{flushleft}

We resort to random matrix theory to solve the above challenge. Specifically,
we first derive deterministic equivalents (DEs) \cite{Wagner_TIT12s_LargeMIMO}
for the conditional average rate and power. Then we obtain an approximated
problem $\mathcal{P}_{E}\left(\mathcal{G}_{T}\left(\mathbf{\Theta}\right)\right)$
by replacing the conditional average rate and power with their DE
approximations. Finally, we show that the solution of $\mathcal{P}_{E}\left(\mathcal{G}_{T}\left(\mathbf{\Theta}\right)\right)$
is an $O\left(\nu\right)$-\textit{optimal solution} of $\mathcal{P}\left(\mathcal{G}_{T}\left(\mathbf{\Theta}\right)\right)$.
\begin{defn}
[$O\left(\nu\right)$-optimal solution]A solution $\Omega=\left\{ \Xi,\mathbf{q}\right\} $
is called an $O\left(\nu\right)$-\textit{feasible solution} of $\mathcal{P}\left(\mathcal{G}_{T}\left(\mathbf{\Theta}\right)\right)$
if it satisfies the zero inter-cell interference constraint $\mathbf{F}_{n}^{\dagger}\left(j\right)\sum_{k\in\overline{\mathcal{S}}_{n}\left(j\right)\triangleq\overline{\mathcal{U}}_{n}\cap\mathcal{S}\left(j\right)}\mathbf{\Theta}_{k,n}=\mathbf{0},\forall n,j$
and the following relaxed per-BS power constraint
\[
\textrm{E}\left[\left.P_{n}\left(\Gamma_{j}\right)\right|\mathbf{\Theta}\right]-P_{c}\leq O\left(\nu\right),\:\forall j.
\]
It is called an $O\left(\nu\right)$-\textit{optimal solution} of
$\mathcal{P}\left(\mathcal{G}_{T}\left(\mathbf{\Theta}\right)\right)$
if it is an $O\left(\nu\right)$-feasible solution and $U^{*}-U\left(\overline{\mathbf{r}}\left(\Omega|\mathbf{\Theta}\right)\right)\leq O\left(\nu\right)$,
where $U^{*}$ is the optimal objective value of $\mathcal{P}\left(\mathcal{G}_{T}\left(\mathbf{\Theta}\right)\right)$.
\end{defn}

Throughout the paper, the notation $M\rightarrow\infty$ refers to
$M\rightarrow\infty$ and $\left|\mathcal{U}_{n}\right|\rightarrow\infty,\forall n$
such that $0<\underset{M\rightarrow\infty}{\liminf}\left|\mathcal{U}_{n}\right|/M\leq\underset{M\rightarrow\infty}{\limsup}\left|\mathcal{U}_{n}\right|/M<\infty$.
For technical reasons, we require the following assumptions.

\begin{assumption}[Technical Assumptions for DE]\label{asm:LSFM}$\:$
\begin{enumerate}
\item All spatial correlation matrices $\mathbf{\Theta}_{k,n},\forall k,n$
have uniformly bounded spectral norm on $M$, i.e., 
\begin{eqnarray}
\underset{M\rightarrow\infty}{\limsup}\underset{1\leq k\leq K}{\textrm{sup}}\left\Vert \mathbf{\Theta}_{k,n}\right\Vert  & < & \infty,\:\forall n.\label{eq:LFbound}
\end{eqnarray}
Moreover, $\underset{M\rightarrow\infty}{\liminf}\frac{1}{M}\textrm{Rank}\left(\sum_{k\in\mathcal{U}_{n}}\mathbf{\Theta}_{k,n}\right)>0$.
\item All the random matrices $\frac{1}{M}\mathbf{H}_{\mathcal{U}_{n}}\mathbf{H}_{\mathcal{U}_{n}}^{\dagger},\forall n$
have uniformly bounded spectral norm on $M$ with probability one,
i.e., 
\begin{eqnarray*}
\underset{M\rightarrow\infty}{\limsup}\left\Vert \frac{1}{M}\mathbf{H}_{\mathcal{U}_{n}}\mathbf{H}_{\mathcal{U}_{n}}^{\dagger}\right\Vert  & \overset{a.s}{<} & \infty,\:\forall n.
\end{eqnarray*}

\item $w_{k}=O\left(1/K\right),\: k=1,...,K$.\hfill \IEEEQED
\end{enumerate}
\end{assumption}

Assumption \ref{asm:LSFM}-1) is satisfied by many MIMO channel models
such as the angular domain MIMO channel model in \cite{Tse_05book_wireless_communication}
and it is a standard assumption in the literatures, see e.g., \cite{Wagner_TIT12s_LargeMIMO,Debbah_JSAC12_MCellRMT}.
Under Assumption \ref{asm:LSFM}-1), Assumption \ref{asm:LSFM}-2)
holds true if $\underset{M\rightarrow\infty}{\limsup}\left|\left\{ \mathbf{\Theta}_{k,n}:\: k\in\mathcal{U}_{n}\right\} \right|<\infty$,
that is, if $\left\{ \mathbf{\Theta}_{k,n}:\: k\in\mathcal{U}_{n}\right\} $
belongs to a finite family \cite{Wagner_TIT12s_LargeMIMO}. According
to the locally-clustered channel model in Assumption \ref{asm:Chmodel},
we have$\left|\left\{ \mathbf{\Theta}_{k,n}:\: k\in\mathcal{U}_{n}\right\} \right|\leq N_{c}$
and thus Assumption \ref{asm:LSFM}-2) holds true. Assumption \ref{asm:LSFM}-3)
is to ensure that the utility function is bounded as $K\rightarrow\infty$. 
\begin{lem}
[DE of Rate and Power]\label{lem:AsyWS}Let Assumption \ref{asm:LSFM}
hold true and consider composite control variable $\Gamma=\left\{ \mathbf{F},\mathcal{S},\mathbf{p}\right\} $
satisfying: 1) $\Gamma\in\Xi^{\textrm{A}}$; 2) the corresponding
user selection $\mathcal{S}_{n}$ satisfies $0<\underset{M\rightarrow\infty}{\liminf}\left|\mathcal{S}_{n}\right|/M\leq\underset{M\rightarrow\infty}{\limsup}\left|\mathcal{S}_{n}\right|/M<\infty$.
Then for any BS $n$, we have
\begin{eqnarray*}
\lim_{M\rightarrow\infty}\left|r_{k}\left(\Gamma\right)-r_{k}^{\circ}\left(\Gamma|\mathbf{\Theta}\right)\right| & \overset{a.s}{\leq} & O\left(\nu\right),\:\forall k\in\mathcal{S}_{n},\\
\lim_{M\rightarrow\infty}\left|P_{n}\left(\Gamma\right)-P_{n}^{\circ}\left(\Gamma|\mathbf{\Theta}\right)\right| & \overset{a.s}{\leq} & O\left(\nu\right),
\end{eqnarray*}
for sufficiently small $\nu>0$, where
\begin{eqnarray}
r_{k}^{\circ}\left(\Gamma|\mathbf{\Theta}\right) & = & \log\left(1+p_{k}\right),\:\forall k\in\mathcal{S}_{n},\label{eq:Asyrate}\\
P_{n}^{\circ}\left(\Gamma|\mathbf{\Theta}\right) & = & \frac{1}{M}\sum_{i\in\mathcal{S}_{n}}\frac{p_{i}}{\xi_{i}},\label{eq:AsyBSpow}
\end{eqnarray}
are the deterministic equivalent (DE) of user rate and BS transmit
power, and $\xi_{i},\forall i\in\mathcal{S}_{n}$ form the unique
solution of
\begin{eqnarray}
\xi_{i} & = & \frac{1}{M}\textrm{Tr}\left(\tilde{\mathbf{\Theta}}_{i,n}\mathbf{T}_{n}\right),\nonumber \\
\mathbf{T}_{n} & = & \left(\frac{1}{M}\sum_{j\in\mathcal{S}_{n}}\frac{\tilde{\mathbf{\Theta}}_{j,n}}{\nu+\xi_{j}}+\mathbf{I}_{M}\right)^{-1},\label{eq:AkfixEqu}
\end{eqnarray}
with $\tilde{\mathbf{\Theta}}_{i,n}=\mathbf{F}_{n}\mathbf{F}_{n}^{\dagger}\mathbf{\Theta}_{i,n}\mathbf{F}_{n}\mathbf{F}_{n}^{\dagger},\forall i\in\mathcal{S}_{n}$.
\end{lem}

Please refer to Appendix \ref{sub:Proof-of-LemmaAsyWS} for the proof. 
\begin{rem}
Note that the above DEs are established on the \textquotedbl{}conditional
distribution of the channel\textquotedbl{} (conditioned on the statistics\textbf{
}\textbf{\textit{$\mathbf{\Theta}$}}). Given a realization of \textit{$\mathbf{\Theta}$}
(the statistics), the control actions $\Gamma=\left\{ \mathbf{F},\mathcal{S},\mathbf{p}\right\} $
are all fixed (because they are adaptive to \textit{$\mathbf{\Theta}$}
only). As such, the conditional measure of $\mathbf{H}$ (conditioned
on the given \textit{$\mathbf{\Theta}$}) will exhibit ``random matrix
theory'' behavior and the DE convergence in Lemma \ref{lem:AsyWS}
can be proved using standard techniques in \cite{Wagner_TIT12s_LargeMIMO}.
On the other hand, if $\mathcal{S}$ were adaptive to the instantaneous
CSI $\mathbf{H}$ (short-term user selection), then conditioned on
$\mathbf{\Theta}$, $\mathcal{S}$ would be random and hence the DE
approximation would fail (due to the random $\mathcal{S}$ or extreme
value effect of the user selection which changes the underlying conditional
distribution of the channels $\mathbf{H}$). Similar conclusion has
also been made in \cite{Caire_TIT13_JSDM} that the DE of the data
rate in\textbf{ }massive MIMO system is valid as long as the user
selection is independent of the instantaneous CSI $\mathbf{H}$.
\end{rem}

Based on Lemma \ref{lem:AsyWS}, we have the following result.
\begin{thm}
[Asymptotic $O\left(\nu\right)$-equivalence of $\mathcal{P}\left(\mathcal{G}_{T}\left(\mathbf{\Theta}\right)\right)$]\label{thm:AsyEqP}Let
$\Omega^{\star}$ denote the optimal solution of 
\begin{eqnarray*}
\mathcal{P}_{E}\left(\mathcal{G}_{T}\left(\mathbf{\Theta}\right)\right):\:\underset{\Omega}{\textrm{max}}\: U_{E}\left(\Omega\right)\triangleq\sum_{k=1}^{K}w_{k}u\left(\sum_{j=1}^{\left|\Xi\right|}q_{j}r_{k}^{\circ}\left(\Gamma_{j}|\mathbf{\Theta}\right)\right)\\
\textrm{s.t.}\:\Omega\in\Lambda^{\circ}\left(P_{c}\right).\;\;\;\;\;\;\;\;\;\;\;\;\;\;\;\;\;\;\;\;\;\;\;\;\;\;\;\;\;\;\;\;\;\;\;\;\;\;\;\;\;
\end{eqnarray*}
where $\Lambda^{\circ}\left(P_{c}\right)=\left\{ \Omega\triangleq\left\{ \Xi,\mathbf{q}\right\} :\:\Xi\subseteq\Xi^{\textrm{F}\circ}\left(P_{c}\right)\right\} $,
and $\Xi^{\textrm{F}\circ}\left(P_{c}\right)=\left\{ \Gamma\in\Xi^{\textrm{A}}:\: P_{n}^{\circ}\left(\Gamma|\mathbf{\Theta}\right)\leq P_{c}\right\} $.
Given Assumption \ref{asm:LSFM} and for sufficiently small $\nu>0$,
$\Omega^{\star}$ is an $O\left(\nu\right)$-optimal solution of $\mathcal{P}\left(\mathcal{G}_{T}\left(\mathbf{\Theta}\right)\right)$
as $M\rightarrow\infty$.\hfill \IEEEQED
\end{thm}

Please refer to Appendix \ref{sub:Proof-of-TheoremAsyEqP} for the
proof. By Theorem \ref{thm:AsyEqP}, the solution of $\mathcal{P}\left(\mathcal{G}_{T}\left(\mathbf{\Theta}\right)\right)$
can be approximated by the solution of $\mathcal{P}_{E}$, and the
approximation is $O\left(\nu\right)$-optimal as $M\rightarrow\infty$.

\section{Solution of $\mathcal{P}_{E}\left(\mathcal{G}_{T}\left(\mathbf{\Theta}\right)\right)$\label{sec:Solution-forPE}}

In the rest of the paper, we focus on solving $\mathcal{P}_{E}\left(\mathcal{G}_{T}\left(\mathbf{\Theta}\right)\right)$
for fixed $\mathbf{\Theta}$. We will use $\mathcal{G}_{T}$, $r_{k}^{\circ}\left(\Gamma\right)$
and $P_{n}^{\circ}\left(\Gamma\right)$ as simplified notations for
$\mathcal{G}_{T}\left(\mathbf{\Theta}\right)$, $r_{k}^{\circ}\left(\Gamma|\mathbf{\Theta}\right)$
and $P_{n}^{\circ}\left(\Gamma|\mathbf{\Theta}\right)$ when there
is no ambiguity. Clearly, the utility function $U_{E}\left(\Omega\right)$
is not a convex function on $\Omega$ and thus $\mathcal{P}_{E}\left(\mathcal{G}_{T}\right)$
is a non-convex optimization problem. Moreover, the optimization variables
in $\mathcal{P}_{E}\left(\mathcal{G}_{T}\right)$ involve a set of
composite control variables $\Xi$ with undetermined size and the
associated probabilities $\mathbf{q}$ with undetermined dimension.
It is in general very difficult to find the global optimal solution
for such a non-convex problem. In this section, we are going to address
the following challenge.\vspace{-3bp}

\begin{flushleft}
\fbox{\begin{minipage}[t]{0.96\columnwidth}%
\begin{challenge}[Design a Global Convergent Algorithm for $\mathcal{P}_{E}\left(\mathcal{G}_{T}\right)$]\label{chl:Non-convex}Exploit
the specific structure of problem $\mathcal{P}_{E}\left(\mathcal{G}_{T}\right)$
to design an iterative algorithm that converges to the global optimal
solution of $\mathcal{P}_{E}\left(\mathcal{G}_{T}\right)$.\end{challenge}%
\end{minipage}}
\par\end{flushleft}

We first study the optimality condition of $\mathcal{P}_{E}\left(\mathcal{G}_{T}\right)$.
Then we propose an iterative algorithm to solve Challenge \ref{chl:Non-convex}.

\subsection{Global Optimality Condition of $\mathcal{P}_{E}\left(\mathcal{G}_{T}\right)$\label{sub:Global-Optimality-Condition}}

It is difficult to find a simple characterization for the necessary
and sufficient global optimality condition of a general non-convex
problem. However, problem \textbf{$\mathcal{P}_{E}\left(\mathcal{G}_{T}\right)$
}is not an arbitrary non-convex problem but has some specific hidden
convexity structure, which can be exploited to derive the global optimality
condition for \textbf{$\mathcal{P}_{E}\left(\mathcal{G}_{T}\right)$}
as shown below.

We first study the hidden convexity of \textbf{$\mathcal{P}_{E}\left(\mathcal{G}_{T}\right)$}.
Define the (deterministic equivalent of) average rate region as: 
\begin{equation}
\mathcal{R}\triangleq\underset{\Omega\in\Lambda^{\circ}\left(P_{c}\right)}{\bigcup}\left\{ \mathbf{x}\in\mathbb{R}_{+}^{K}:\:\mathbf{x}\le\overline{\mathbf{r}}^{\circ}\left(\Omega\right)\right\} ,\label{eq:RegionR}
\end{equation}
where $\overline{\mathbf{r}}^{\circ}\left(\Omega\right)=\left[\overline{r}_{1}^{\circ}\left(\Omega\right),...,\overline{r}_{K}^{\circ}\left(\Omega\right)\right]^{T}$
with $\overline{r}_{k}^{\circ}\left(\Omega\right)=\sum_{j=1}^{\left|\Xi\right|}q_{j}r_{k}^{\circ}\left(\Gamma_{j}\right)$.
Then we have the following Lemma.
\begin{lem}
[Convexity of $\mathcal{R}$]\label{lem:Convexity-of-R}$\mathcal{R}=\textrm{Conv}\left(\mathcal{R}^{\textrm{F}}\right)$,
where $\textrm{Conv}\left(\cdot\right)$ denotes the convex hull operation
and $\mathcal{R}^{\textrm{F}}\triangleq\left\{ \mathbf{r}^{\circ}\left(\Gamma\right)=\left[r_{1}^{\circ}\left(\Gamma\right),...,r_{K}^{\circ}\left(\Gamma\right)\right]^{T}:\Gamma\in\Xi^{\textrm{F}\circ}\left(P_{c}\right)\right\} $.
\end{lem}

Please refer to Appendix \ref{sub:Proofs-for-the-GOP} for the proof. 

The following lemma shows that problem $\mathcal{P}_{E}\left(\mathcal{G}_{T}\right)$
is equivalent to a convex problem:
\begin{equation}
\underset{\overline{\mathbf{r}}^{\circ}}{\textrm{max}}U\left(\overline{\mathbf{r}}^{\circ}\right)\triangleq\sum_{k=1}^{K}w_{k}u\left(\overline{r}_{k}^{\circ}\right),\:\textrm{s.t.}\:\overline{\mathbf{r}}^{\circ}=\left[\overline{r}_{1}^{\circ},...,\overline{r}_{K}^{\circ}\right]^{T}\in\mathcal{R},\label{eq:EquPLApre-1}
\end{equation}

\begin{lem}
[Equivalence between $\mathcal{P}_{E}\left(\mathcal{G}_{T}\right)$
and (\ref{eq:EquPLApre-1})]\label{lem:Hiddencvx}If $\Omega^{\star}$
is the global optimal solution of $\mathcal{P}_{E}\left(\mathcal{G}_{T}\right)$,
then $\overline{\mathbf{r}}^{\circ}\left(\Omega^{\star}\right)$ is
the optimal solution of problem (\ref{eq:EquPLApre-1}); on the other
hand, if $\overline{\mathbf{r}}^{\circ\star}$ is the optimal solution
of problem (\ref{eq:EquPLApre-1}), then any $\Omega^{\star}$ satisfying
$\overline{\mathbf{r}}^{\circ}\left(\Omega^{\star}\right)=\overline{\mathbf{r}}^{\circ\star}$
is also the global optimal solution of $\mathcal{P}_{E}\left(\mathcal{G}_{T}\right)$. 
\end{lem}

Please refer to Appendix \ref{sub:Proofs-for-the-GOP} for the proof.
This hidden convexity of $\mathcal{P}_{E}\left(\mathcal{G}_{T}\right)$
(i.e., the equivalence between $\mathcal{P}_{E}\left(\mathcal{G}_{T}\right)$
and (\ref{eq:EquPLApre-1})) is the key to derive the global optimality
condition of $\mathcal{P}_{E}\left(\mathcal{G}_{T}\right)$. Note
that although problem (\ref{eq:EquPLApre-1}) is convex, the solution
is still non-trivial because there is no simple characterization for
its feasible set $\mathcal{R}$.

To derive the global optimality condition of $\mathcal{P}_{E}\left(\mathcal{G}_{T}\right)$,
we also need the first order optimality condition of problem (\ref{eq:EquPLApre-1})
as summarized in the following lemma.
\begin{lem}
[First Order Optimality Condition of (\ref{eq:EquPLApre-1})]\label{lem:First-Order-OptimalityHD}A
solution $\overline{\mathbf{r}}^{\circ\star}=\left[\overline{r}_{1}^{\circ\star},...,\overline{r}_{K}^{\circ\star}\right]^{T}$
is optimal for problem (\ref{eq:EquPLApre-1}) if and only if
\begin{equation}
\nabla^{T}U\left(\overline{\mathbf{r}}^{\circ\star}\right)\left(\overline{\mathbf{r}}^{\circ\star}-\mathbf{x}\right)\geq0,\forall\mathbf{x}\in\mathcal{R}.\label{eq:OptCondHD}
\end{equation}

\end{lem}

Finally, from Lemma \ref{lem:Hiddencvx} and Lemma \ref{lem:First-Order-OptimalityHD},
we can obtain the necessary and sufficient global optimality condition
for problem $\mathcal{P}_{E}\left(\mathcal{G}_{T}\right)$ as follows.
\begin{thm}
[Global Optimality Condition of $\mathcal{P}_{E}\left(\mathcal{G}_{T}\right)$]\label{thm:Global-Optimality-ConditionPE}A
control policy $\Omega^{\star}=\left\{ \Xi^{\star},\mathbf{q}^{\star}\right\} $
with $\Xi^{\star}=\left\{ \Gamma_{1}^{\star},...,\Gamma_{\left|\Xi^{\star}\right|}^{\star}\right\} $
is a global optimal solution of $\mathcal{P}_{E}\left(\mathcal{G}_{T}\right)$
if and only if $\Gamma_{j}^{\star},\forall j\in\left\{ 1,...,\left|\Xi^{\star}\right|\right\} $
satisfies:
\begin{equation}
\boldsymbol{\mu}^{\star T}\left(\mathbf{r}^{\circ}\left(\Gamma_{j}^{\star}\right)-\mathbf{r}^{\circ}\left(\Gamma\right)\right)\geq0,\:\forall\Gamma\in\Xi^{\textrm{F}\circ}\left(P_{c}\right),\label{eq:OptcondPE}
\end{equation}
where $\mathbf{r}^{\circ}\left(\Gamma\right)=\left[r_{1}^{\circ}\left(\Gamma\right),...,r_{K}^{\circ}\left(\Gamma\right)\right]^{T}$
and the weight vector $\boldsymbol{\mu}^{\star}\triangleq\nabla U\left(\overline{\mathbf{r}}^{\circ}\left(\Omega^{\star}\right)\right)=\left[w_{k}\frac{\partial u\left(r\right)}{\partial r}|_{r=\overline{r}_{k}^{\circ}\left(\Omega^{\star}\right)}\right]_{k=1,...,K}$. 
\end{thm}

The detailed proof can be found in Appendix \ref{sub:Proofs-for-the-GOP}.

\subsection{Global Optimal Solution of $\mathcal{P}_{E}\left(\mathcal{G}_{T}\right)$\label{sub:Global-Optimal-Solution}}

Just as we can obtain the optimal solution of a convex problem by
solving its KKT conditions, we can also obtain the global optimal
solution of $\mathcal{P}_{E}\left(\mathcal{G}_{T}\right)$ by solving
the global optimality condition in Theorem \ref{thm:Global-Optimality-ConditionPE}.
Specifically, for any given spatial correlation matrices $\mathbf{\Theta}$,
we propose Algorithm E to achieve the global optimality condition
of $\mathcal{P}_{E}\left(\mathcal{G}_{T}\left(\mathbf{\Theta}\right)\right)$
by iteratively updates the optimization variables $\Xi,\mathbf{q}$
and the weight vector\textit{ }$\boldsymbol{\mu}$ in Theorem \ref{thm:Global-Optimality-ConditionPE}. 

\smallskip{}

\textit{Algorithm E} (Top level algorithm for solving $\mathcal{P}_{E}\left(\mathcal{G}_{T}\left(\mathbf{\Theta}\right)\right)$): 

\textbf{\small{}Initialization}{\small{}: Set $i=0$ and let $\boldsymbol{\mu}^{(0)}=\left[w_{k}\right]_{k=1,...,K}$.
Call Procedure $\textrm{W}^{\star}$ with input $\boldsymbol{\mu}^{(0)}$
to obtain a composite control variable $\Gamma^{\star}\left(\boldsymbol{\mu}^{(0)}\right)$
and let $\Xi^{(0)}=\left\{ \Gamma^{\star}\left(\boldsymbol{\mu}^{(0)}\right)\right\} $.}{\small \par}

\textbf{\small{}Step 1 }{\small{}(Update probability vector $\mathbf{q}$):}\textbf{\small{}
}{\small{}Call Procedure Q with input $\Xi^{(i)}=\left\{ \Gamma_{1}^{(i)},...,\Gamma_{\left|\Xi^{(i)}\right|}^{(i)}\right\} $
to obtain the updated probability vector $\mathbf{q}^{(i)}=\left[q_{j}^{(i)}\right]_{j=1,...,\left|\Xi^{(i)}\right|}$.
Let $\widetilde{\Xi}^{(i)}=\left\{ \Gamma_{j}^{(i)}:\: j\in\mathcal{J}^{(i)}\right\} $
and $\widetilde{\mathbf{q}}^{(i)}=\left[q_{j}^{(i)}\right]_{j\in\mathcal{J}^{(i)}}$,
where $\mathcal{J}^{(i)}=\left\{ j:\: q_{j}^{(i)}>0\right\} $. Let
$\Omega^{(i)}=\left\{ \widetilde{\Xi}^{(i)},\widetilde{\mathbf{q}}^{(i)}\right\} $}{\small \par}

\textbf{\small{}Step 2}{\small{} (Update composite control variable
set $\Xi$): Let 
\begin{equation}
\boldsymbol{\mu}^{(i+1)}=\nabla U\left(\overline{\mathbf{r}}^{\circ}\left(\Omega^{(i)}\right)\right)\triangleq\left[w_{k}\frac{\partial u\left(r\right)}{\partial r}|_{r=\overline{r}_{k}^{\circ}\left(\Omega^{(i)}\right)}\right]_{k=1,...,K}.\label{eq:mui}
\end{equation}
Call Procedure $\textrm{W}^{\star}$ with input $\boldsymbol{\mu}^{(i+1)}$
to obtain a new composite control variable $\Gamma^{\star}\left(\boldsymbol{\mu}^{(i+1)}\right)$.
Update $\Xi$ as 
\begin{equation}
\Xi^{(i+1)}=\widetilde{\Xi}^{(i)}\cup\Gamma^{\star}\left(\boldsymbol{\mu}^{(i+1)}\right).\label{eq:SBset}
\end{equation}
}{\small \par}

\textbf{\small{}Step 3}{\small{}: If $i>0$}\textbf{\small{} }{\small{}and}\textbf{\small{}
$\left|U_{E}\left(\Omega^{(i)}\right)-U_{E}\left(\Omega^{(i-1)}\right)\right|\leq\varepsilon$}{\small{},
where $\varepsilon>0$ is a small number, terminate the algorithm.
Otherwise, let}\textbf{\small{} $i=i+1$ }{\small{}and return to Step
1.}{\small \par}

\smallskip{}

\begin{figure}
\begin{centering}
\textsf{\includegraphics[clip,width=90mm]{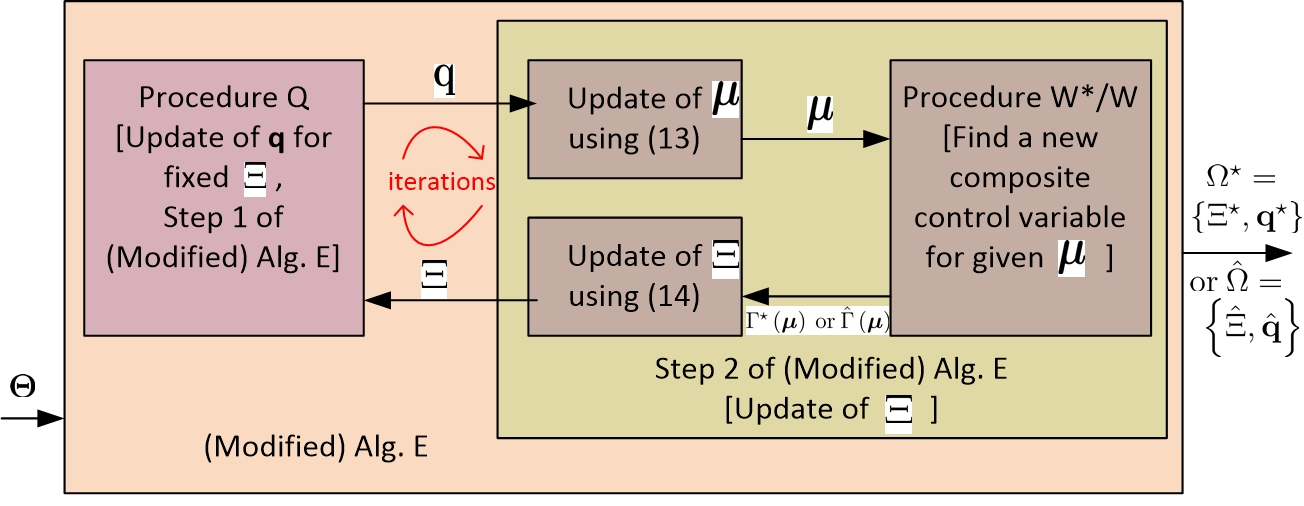}}
\par\end{centering}

\protect\caption{\label{fig:Alg_connec}{\small{}Summary of overall solution and the
inter-relationship of the algorithm components for both Algorithm
E (with Procedure $\textrm{W}^{\star}$, composite control variable
$\Gamma^{\star}\left(\boldsymbol{\mu}\right)$ and output $\Omega^{\star}=\left\{ \Xi^{\star},\mathbf{q}^{\star}\right\} $)
and the modified Algorithm E (with Procedure W, composite control
variable $\hat{\Gamma}\left(\boldsymbol{\mu}\right)$ and output $\hat{\Omega}=\left\{ \hat{\Xi},\hat{\mathbf{q}}\right\} $).
The iteration number $^{(i)}$ is omitted for simplicity. Each square
represents an algorithm component and the corresponding square bracket
explains the function of this algorithm component.}}
\end{figure}

Fig. \ref{fig:Alg_connec} summarizes the inter-relationship between
the components of Algorithm E. Algorithm E contains two procedures
(subroutines) which will be elaborated below.
\begin{rem}
Algorithm E can be interpreted as the Frank-Wolfe Algorithm (also
known as the conditional gradient algorithm) with exact line search
\cite{Frank_NRLQ56_condgradient} applied on the equivalent convex
problem in (\ref{eq:EquPLApre-1}). Compared to the conventional Frank-Wolfe
Algorithm, the main difference is that the optimization variable in
problem $\mathcal{P}_{E}\left(\mathcal{G}_{T}\right)$ is $\Omega$
instead of $\overline{\mathbf{r}}^{\circ}$ in (\ref{eq:EquPLApre-1}),
and the optimization w.r.t. $\Omega$ is non-convex. Nonetheless,
we can exploit the hidden convexity (global optimality condition)
in Theorem \ref{thm:Global-Optimality-ConditionPE} to establish the
global convergence of Algorithm E as will be shown in Theorem \ref{thm:OptAE}.
\end{rem}

\subsubsection{Procedure Q (Optimization of $\mathbf{q}$ for fixed $\Xi$)}

For given input $\Xi$, Procedure Q essentially solves the optimal
probability vector $\mathbf{q}$ for $\mathcal{P}_{E}\left(\mathcal{G}_{T}\right)$
with fixed $\Xi$, i.e., Procedure Q with input $\Xi$ is a standard
convex optimization procedure to solve the following optimization
problem:
\begin{eqnarray}
 & \underset{\left[q_{j}\right]_{j=1,...,\left|\Xi\right|}}{\max}\sum_{k=1}^{K}w_{k}u\left(\sum_{j=1}^{\left|\Xi\right|}q_{j}r_{k}^{\circ}\left(\Gamma_{j}\right)\right),\label{eq:fixthetaPdet}\\
 & \textrm{s.t.}\: q_{j}\in\left[0,1\right],\forall j\:\textrm{and}\:\sum_{j=1}^{\left|\Xi\right|}q_{j}=1,\nonumber 
\end{eqnarray}
where $\Gamma_{j}$ is the $j$-th composite control variable in $\Xi$.
Hence, Procedure Q can be efficiently implemented by existing convex
optimization methods/software. As such, the pseudo code of Procedure
Q is omitted here for conciseness.

\subsubsection{Procedure $\textrm{W}^{\star}$ (Finding a new composite control
variable for given $\boldsymbol{\mu}$)}

The pseudo code of Procedure $\textrm{W}^{\star}$ is summarized in
Table \ref{tab:AlgES}. In Line 2, $\xi_{i},\forall i\in\mathcal{S}_{n}$
is the unique solution of (\ref{eq:AkfixEqu}) with $\tilde{\mathbf{\Theta}}_{i,n}=\tilde{\mathbf{\Theta}}_{i,n}\left(\overline{\mathcal{S}}_{n}\right)\triangleq\left(\mathbf{I}_{M}-\mathbf{U}\left(\overline{\mathcal{S}}_{n}\right)\mathbf{U}^{\dagger}\left(\overline{\mathcal{S}}_{n}\right)\right)\mathbf{\Theta}_{i,n}\left(\mathbf{I}_{M}-\mathbf{U}\left(\overline{\mathcal{S}}_{n}\right)\mathbf{U}^{\dagger}\left(\overline{\mathcal{S}}_{n}\right)\right)$,
where $\mathbf{U}\left(\overline{\mathcal{S}}_{n}\right)=\textrm{orth}\left(\sum_{k\in\overline{\mathcal{S}}_{n}}\mathbf{\Theta}_{k,n}\right)$.
In Line 4, $R\left(\mathcal{S}\right)$ is the (deterministic equivalent
of) weighted sum-rate for given user selection $\mathcal{S}$. In
Line 7, $\overline{\mathcal{S}}_{n}^{\star}=\overline{\mathcal{U}}_{n}\cap\mathcal{S}^{\star}$.
For convenience, $\xi_{k}$ is referred to as the \textit{effective
channel gain} of user $k$ and $\tilde{\mathbf{\Theta}}_{i,n}\left(\overline{\mathcal{S}}_{n}\right)$
is called the \textit{projected spatial correlation matrix} of user
$i$. For conciseness, $\tilde{\mathbf{\Theta}}_{i,n}\left(\overline{\mathcal{S}}_{n}\right)$
is denoted as $\tilde{\mathbf{\Theta}}_{i,n}$ when there is no ambiguity.
To calculate the weighted sum-rate $R\left(\mathcal{S}\right)$, we
need to obtain the effective channel gains $\xi_{i},\forall i\in\mathcal{S}$
associated with $\mathcal{S}$ by solving the fixed point equation
in (\ref{eq:AkfixEqu}). The solution of (\ref{eq:AkfixEqu}) can
be obtained using the following fixed point iterations \cite{Wagner_TIT12s_LargeMIMO}
\begin{equation}
\xi_{i}^{(t+1)}=\frac{1}{M}\textrm{Tr}\left(\tilde{\mathbf{\Theta}}_{i,n}\left(\frac{1}{M}\sum_{j\in\mathcal{S}_{n}}\frac{\tilde{\mathbf{\Theta}}_{j,n}}{\nu+\xi_{j}^{(t)}}+\mathbf{I}_{M}\right)^{-1}\right),\label{eq:Fixiter}
\end{equation}
with initial point $\xi_{i}^{(0)}=1,\forall i\in\mathcal{S}_{n}$,
where $\tilde{\mathbf{\Theta}}_{i,n}=\tilde{\mathbf{\Theta}}_{i,n}\left(\overline{\mathcal{S}}_{n}\right),\forall i\in\mathcal{S}_{n}$.

\begin{table}
\protect\caption{\label{tab:AlgES}Procedure $\textrm{W}^{\star}$ (for solving Condition
(\ref{eq:OptcondPE-1}))}

\centering{}%
\begin{tabular}{l}
\hline 
{\small{}1.}\textbf{\small{} For }{\small{}all $\mathcal{S}$}\tabularnewline
{\small{}2.$\;$$\;$$\;$$\;$Let}\textbf{\small{} }{\small{}$p_{k}^{\star}\left(\mathcal{S}\right)=\left(\frac{\mu_{k}M\xi_{k}}{\lambda_{b_{k}}}-1\right)^{+},\forall k$,
where $\lambda_{b_{k}}$ is }\tabularnewline
{\small{}3.$\;$$\;$$\;$$\;$chosen such that $\frac{1}{M}\sum_{i\in\mathcal{S}_{b_{k}}}\frac{p_{i}^{\star}\left(\mathcal{S}\right)}{\xi_{i}}=P_{c}$.}\tabularnewline
{\small{}4.$\;$$\;$$\;$$\;$Let $R\left(\mathcal{S}\right)=\sum_{k\in\mathcal{S}}\mu_{k}\textrm{log}\left(1+p_{k}^{\star}\left(\mathcal{S}\right)\right)$.}\tabularnewline
{\small{}5.}\textbf{\small{} End}\tabularnewline
{\small{}6. Let $\mathcal{S}^{\star}=\underset{\mathcal{S}}{\textrm{argmax}}R\left(\mathcal{S}\right)$
and $\mathbf{p}^{\star}=\left[p_{k}^{\star}\left(\mathcal{S}^{\star}\right)\right]_{k\in\mathcal{S}^{\star}}$.}\tabularnewline
{\small{}7. Let}\textbf{\small{} }{\small{}$\mathbf{F}_{n}^{\star}=\textrm{orth}\left(\left(\mathbf{I}_{M}-\mathbf{U}\left(\overline{\mathcal{S}}_{n}^{\star}\right)\mathbf{U}^{\dagger}\left(\overline{\mathcal{S}}_{n}^{\star}\right)\right)\sum_{k\in\mathcal{S}_{n}^{\star}}\mathbf{\Theta}_{k,n}\right)$.}\tabularnewline
{\small{}8.}\textbf{\small{} Output $\Gamma^{\star}\left(\boldsymbol{\mu}\right)=\left\{ \mathbf{F}^{\star},\mathcal{S}^{\star},\mathbf{p}^{\star}\right\} $}{\small{},
where }\textbf{\small{}$\mathbf{F}^{\star}=\left\{ \mathbf{F}_{1}^{\star},...,\mathbf{F}_{N}^{\star}\right\} $}{\small{}.}\tabularnewline
\hline 
\end{tabular}
\end{table}

For given input $\boldsymbol{\mu}$, Procedure $\textrm{W}^{\star}$
essentially finds a composite control variable $\Gamma^{\star}\left(\boldsymbol{\mu}\right)$
which satisfies the global optimality condition in (\ref{eq:OptcondPE})
for fixed $\boldsymbol{\mu}$.
\begin{thm}
[Characterization of Procedure $\textrm{W}^{\star}$]\label{thm:Char-of-PW}For
given input $\boldsymbol{\mu}$, the output $\Gamma^{\star}\left(\boldsymbol{\mu}\right)$
of Procedure $\textrm{W}^{\star}$ satisfies
\begin{equation}
\boldsymbol{\mu}^{T}\left(\mathbf{r}^{\circ}\left(\Gamma^{\star}\left(\boldsymbol{\mu}\right)\right)-\mathbf{r}_{k}^{\circ}\left(\Gamma\right)\right)\geq0,\:\forall\Gamma\in\Xi^{\textrm{F}\circ}\left(P_{c}\right).\label{eq:OptcondPE-1}
\end{equation}

\end{thm}

Please refer to Appendix \ref{sub:Proof-of-TheoremCharPW} for the
proof.

\subsubsection{Convergence and Performance of Algorithm E\label{sub:Convergence-and-PerformanceE}}

The update rule in Algorithm E is designed according to the global
optimality condition in Theorem 2. As a result, it can be shown that
Algorithm E converges to the global optimal solution of $\mathcal{P}_{E}\left(\mathcal{G}_{T}\right)$
using the global optimality condition in Theorem \ref{thm:Global-Optimality-ConditionPE}
and the property of Algorithm E in the following Lemma.
\begin{lem}
[Property of Algorithm E]\label{lem:Property-of-AlgorithmE}Let
$\Omega^{(i)}$ be the control policy in the $i$-th iteration of
Algorithm E. We have 
\begin{eqnarray}
U_{E}\left(\Omega^{(i+1)}\right)\geq\nonumber \\
\max_{\eta\in\left[0,1\right]}U\left(\left(1-\eta\right)\overline{\mathbf{r}}^{\circ}\left(\Omega^{(i)}\right)+\eta\mathbf{r}^{\circ}\left(\Gamma^{\star}\left(\boldsymbol{\mu}^{(i+1)}\right)\right)\right),\label{eq:maxetaU}
\end{eqnarray}
where $\boldsymbol{\mu}^{(i+1)}=\nabla U\left(\overline{\mathbf{r}}^{\circ}\left(\Omega^{(i)}\right)\right)$
is given in (\ref{eq:mui}).
\end{lem}

Please refer to Appendix \ref{sub:Proofs-for-theOptAE} for the proof.

Using Theorem \ref{thm:Global-Optimality-ConditionPE} and Lemma \ref{lem:Property-of-AlgorithmE},
we obtain the following global convergence result. 
\begin{thm}
[Global Optimality of Algorithm E]\label{thm:OptAE}Algorithm E
monotonically increases the utility $U_{E}\left(\Omega^{(i)}\right)$
and $\lim_{i\rightarrow\infty}U_{E}\left(\Omega^{(i)}\right)\rightarrow U^{\star}$,
where $U^{\star}$ is the global optimal value of $\mathcal{P}_{E}\left(\mathcal{G}_{T}\right)$. 
\end{thm}

Please refer to Appendix \ref{sub:Proofs-for-theOptAE} for the proof.

In step 2 of Algorithm E, we need to call Procedure $\textrm{W}^{\star}$,
which involves an exhaustive user selection process where $R\left(\mathcal{S}\right)$
is calculated for all possible user set $\mathcal{S}$ (see Line 1
to Line 6 of Procedure $\textrm{W}^{\star}$). The complexity of exhaustive
user selection is exponential w.r.t. the number of users $K$. In
the next subsection, we will propose a low complexity solution, named
the\textit{ modified Algorithm E}, for $\mathcal{P}_{E}\left(\mathcal{G}_{T}\right)$
by replacing the exhaustive user selection process with a \textit{statistical
greedy user selection }process\textit{.}

\subsection{Low Complexity Solution of $\mathcal{P}_{E}\left(\mathcal{G}_{T}\right)$\label{sec:Solution-for-WSRNM}}

The low complexity solution (modified Algorithm E) is obtained by
replacing the exact solution $\Gamma^{\star}\left(\boldsymbol{\mu}\right)$
of (\ref{eq:OptcondPE-1}) in step 2 (and the initialization step)
of Algorithm E with an approximate solution $\hat{\Gamma}\left(\boldsymbol{\mu}\right)$
found by a low complexity procedure named \textit{Procedure W}. In
other words, the modified Algorithm E are the same as Algorithm E
except that Procedure $\textrm{W}^{\star}$ (which involves exhaustive
user selection) is replaced by the low complexity counterpart Procedure
W (which is based on statistical greedy user selection). 

The pseudo code of Procedure W is summarized in Table \ref{tab:AlgSGUS}.
In Line 3 and 4, the weighted sum-rate $R\left(\mathcal{S}\right)$
for any given $\mathcal{S}$ can be calculated using the same method
as described in Procedure $\textrm{W}^{\star}$. Clearly, the statistical
greedy user selection loop between Line 2 and Line 9 converges to
a solution $\hat{\mathcal{S}}$ within $K$ iterations. 

\begin{table}
\protect\caption{\label{tab:AlgSGUS}Procedure W (for solving Condition (\ref{eq:OptcondPE-1}))}

\centering{}%
\begin{tabular}{l}
\hline 
{\small{}1.}\textbf{\small{} Initialization}{\small{}: Let $\mathcal{S}=\emptyset$
and $\textrm{Add\_flag}=1$.}\tabularnewline
{\small{}2.$\;$$\;$}\textbf{\small{}while }{\small{}$\textrm{Add\_flag}==1$
and $\left|\mathcal{S}\right|<K$}\tabularnewline
{\small{}3.$\;$$\;$$\;$$\;$Let}\textbf{\small{} }{\small{}$\mathcal{S}^{\circ}=\textrm{argmax}{}_{\mathcal{S}^{'}\in\left\{ \mathcal{S}\cup\left\{ k\right\} :\:\forall k\in\mathcal{U}\backslash\mathcal{S}\right\} }\: R\left(\mathcal{S}^{'}\right)$.}\tabularnewline
{\small{}4.$\;$$\;$$\;$$\;$}\textbf{\small{}if}{\small{} $R\left(\mathcal{S}^{\circ}\right)>R\left(\mathcal{S}\right)$
}\textbf{\small{}then}{\small{} }\tabularnewline
{\small{}5.$\;$$\;$$\;$$\;$$\;$$\;$$\mathcal{S}=\mathcal{S}^{\circ}$. }\tabularnewline
{\small{}6.$\;$$\;$$\;$$\;$}\textbf{\small{}else}\tabularnewline
{\small{}7.$\;$$\;$$\;$$\;$$\;$$\;$$\textrm{Add\_flag}=0$.}\tabularnewline
{\small{}8.$\;$$\;$$\;$$\;$}\textbf{\small{}end if}\tabularnewline
{\small{}9.$\;$$\;$}\textbf{\small{}end while}\tabularnewline
{\small{}10. Let $\hat{\mathcal{S}}=\mathcal{S}$ and $\hat{\mathbf{p}}=\left[p_{k}^{\star}\left(\mathcal{S}\right)\right]_{k\in\mathcal{S}}$.}\tabularnewline
{\small{}11. Let}\textbf{\small{} }{\small{}$\hat{\mathbf{F}}_{n}=\textrm{orth}\left(\left(\mathbf{I}_{M}-\mathbf{U}\left(\overline{\mathcal{S}}_{n}\right)\mathbf{U}^{\dagger}\left(\overline{\mathcal{S}}_{n}\right)\right)\sum_{k\in\mathcal{S}_{n}}\mathbf{\Theta}_{k,n}\right)$.}\tabularnewline
{\small{}12.}\textbf{\small{} Output $\hat{\Gamma}\left(\boldsymbol{\mu}\right)=\left\{ \hat{\mathbf{F}},\hat{\mathcal{S}},\hat{\mathbf{p}}\right\} $}{\small{},
where }\textbf{\small{}$\hat{\mathbf{F}}=\left\{ \hat{\mathbf{F}}_{1},...,\hat{\mathbf{F}}_{N}\right\} $}{\small{}.}\tabularnewline
\hline 
\end{tabular}
\end{table}

Fig. \ref{fig:Alg_connec} summarizes the overall low complexity solution
and the inter-relationship between the components of the modified
Algorithm E. To justify the modified Algorithm E, we need to address
the following challenge.\vspace{-3bp}

\begin{flushleft}
\fbox{\begin{minipage}[t]{0.96\columnwidth}%
\begin{challenge}[Monotone Convergence of the modified Algorithm
E]\label{chl:SGUS}Prove the monotone convergence of the modified
Algorithm E as well as characterize the performance loss of the modified
Algorithm E w.r.t. the global optimal solution.\end{challenge}%
\end{minipage}}
\par\end{flushleft}

The following theorem provides a solution to Challenge \ref{chl:SGUS}.
\begin{thm}
[Convergence of the Modified Alg. E]\label{thm:conve_greedyAE}The
modified Algorithm E monotonically increases the utility $U_{E}\left(\Omega^{(i)}\right)$
and $\lim_{i\rightarrow\infty}U_{E}\left(\Omega^{(i)}\right)\rightarrow\hat{U}$.
Moreover, the gap of $\hat{U}$ with the optimal utility $U_{E}^{\star}$
of $\mathcal{P}_{E}\left(\mathcal{G}_{T}\right)$ is bounded by 
\[
U_{E}^{\star}-\hat{U}\leq\hat{\boldsymbol{\mu}}^{T}\left(\mathbf{r}_{k}^{\circ}\left(\Gamma^{\star}\left(\hat{\boldsymbol{\mu}}\right)\right)-\mathbf{r}_{k}^{\circ}\left(\hat{\Gamma}\left(\hat{\boldsymbol{\mu}}\right)\right)\right),
\]
where $\left(\hat{\overline{\mathbf{r}}}_{k}^{\circ},\hat{\boldsymbol{\mu}}\right)$
can be any accumulation point of the iterates $\left\{ \overline{\mathbf{r}}^{\circ}\left(\Omega^{(i)}\right),\boldsymbol{\mu}^{(i+1)}\right\} $
generated by the modified Algorithm E, $\Gamma^{\star}\left(\hat{\boldsymbol{\mu}}\right)$
is the output of Procedure $\textrm{W}^{\star}$ with input $\hat{\boldsymbol{\mu}}$
and $\hat{\Gamma}\left(\hat{\boldsymbol{\mu}}\right)$ is the output
of Procedure W with input $\hat{\boldsymbol{\mu}}$.
\end{thm}

Please refer to Appendix \ref{sub:Proof-of-TheoremgreedyAE} for the
proof. Theorem \ref{thm:conve_greedyAE} states that the performance
gap between the modified Algorithm E and (the optimal) Algorithm E
is upper bounded by the performance gap (in terms of weighted sum-rate)
between Procedure W (statistical user selection) and Procedure $\textrm{W}^{\star}$
(exhaustive user selection).

\subsubsection*{Complexity Analysis for the Modified Algorithm E}

The computation complexity is evaluated in terms of the number of
matrix multiplications, matrix inversions and Gram\textendash Schmidt
processes, since these operations dominate the first order of the
overall computation complexity. For simplicity, we assume $\textrm{Rank}\left(\mathbf{\Theta}_{i,n}\right)=d,\forall i,n$
and $\left|\overline{\mathcal{U}}_{n}\right|=\overline{K},\forall n$.
Suppose that the fixed point iterations in (\ref{eq:Fixiter}) converges
to the desired accuracy in $C_{f}$ iterations. Then the complexity
of Procedure W is analyzed as follows. In each iteration, the greedy
search to find the $\mathcal{S}^{\circ}$ requires evaluating $K-\left|\mathcal{S}\right|<K$
weighted sum-rates $R\left(\mathcal{S}^{'}\right)$. Each $R\left(\mathcal{S}^{'}\right)$
needs no more than $O\left(C_{f}\left|\mathcal{S}\right|\right)$
$M\times M$ matrix multiplications and $O\left(NC_{f}\right)$ $M\times M$
matrix inversions. The greedy search is repeated for at most $K$
times. For each $n$, $\tilde{\mathbf{\Theta}}_{i,n},\forall i\in\mathcal{U}_{n}$
is updated for at most $K\overline{K}$ times, and each update needs
one $M\times M$ Gram\textendash Schmidt processes (i.e., Gram\textendash Schmidt
processes for a $M\times M$ matrix) to calculate $\textrm{orth}\left(\sum_{j\in\overline{\mathcal{S}}_{n}^{'}}\mathbf{\Theta}_{j,n}\right)$.
Hence, the overall complexity of Procedure W is upper bounded by $O\left(K^{3}C_{f}\right)$
$M\times M$ matrix multiplications, $O\left(K^{2}NC_{f}\right)$
$M\times M$ matrix inversions and $O\left(NK\overline{K}\right)$
$M\times M$ Gram\textendash Schmidt processes. This is also the order
of the per iteration complexity for the modified Algorithm E because
in each iteration of the modified Algorithm E, the computation complexity
is dominated by Procedure W.

\section{Simulation Results\label{sec:Simulation-Results}}

\begin{figure}
\begin{centering}
\includegraphics[width=55mm]{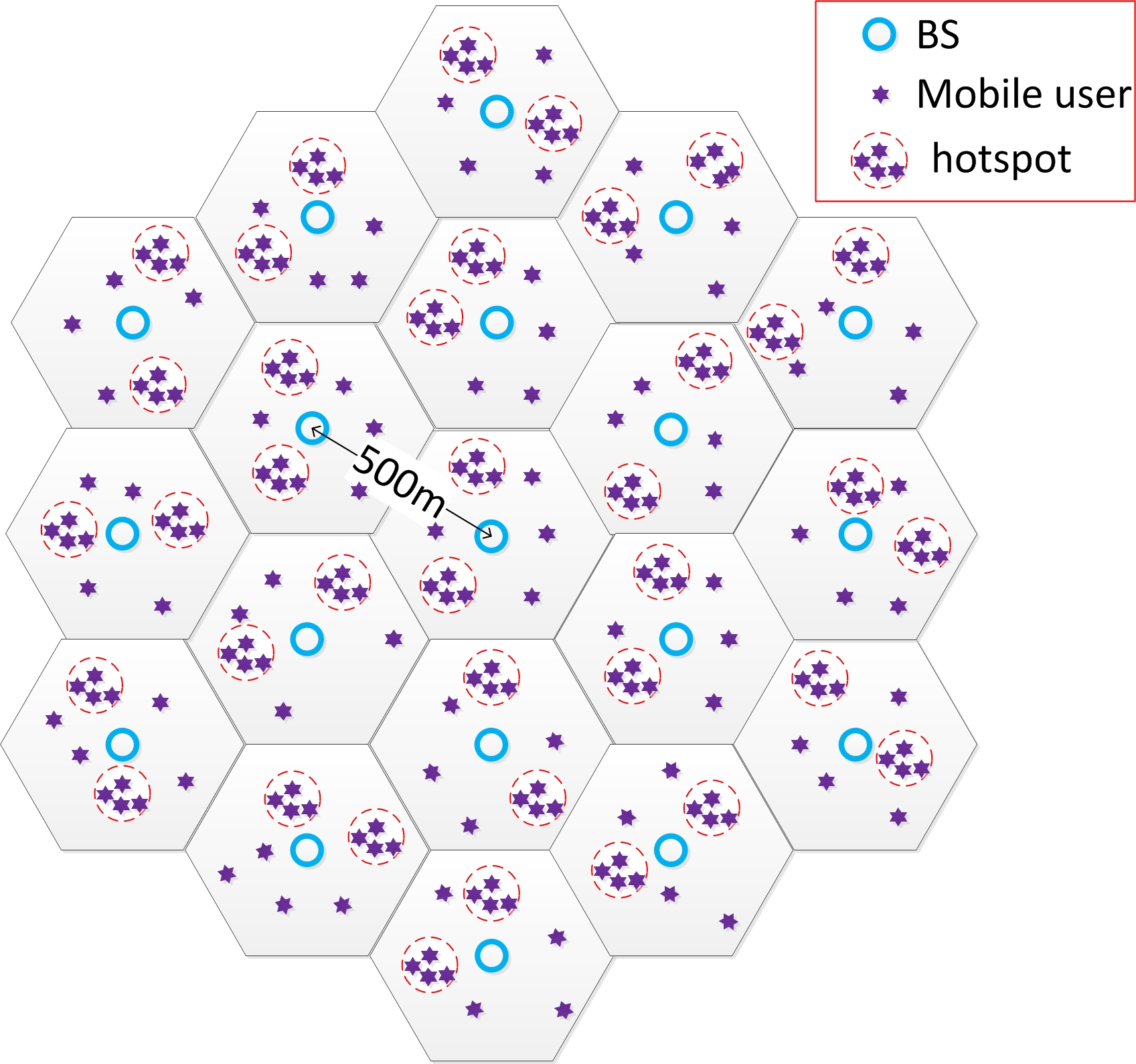}
\par\end{centering}

\protect\caption{\label{fig:HetNet-topo}{\small{}Topology of a cellular network with
$19$ cells.}}
\end{figure}

Consider a cellular network with $19$ cells as illustrated in Fig.
\ref{fig:HetNet-topo}. The inter-site distance is $500$m. In each
cell, there are $2$ uniformly distributed hotspots with a radius
of $50$m. There are $12$ users in one cell, $2/3$ of whom are clustered
around the hotspots, while the others are uniformly distributed within
the cell. Each BS is equipped with $M=48$ antennas. The spatial correlation
matrices are generated according to $\mathbf{\Theta}_{k,n}=L_{k,n}\check{\mathbf{\Theta}}_{k,n},\forall k,n$,
where the path gains $L_{k,n}$'s are generated using the path loss
model (\textquotedblleft Urban Macro NLOS\textquotedblright{} model)
in \cite{3gpp_Rel9}, and the normalized spatial correlation matrices
$\check{\mathbf{\Theta}}_{k,n}$'s with $\textrm{Tr}\left(\check{\mathbf{\Theta}}_{k,n}\right)=M$
and $\textrm{Rank}\left(\mathbf{\Theta}_{k,n}\right)=6$ are randomly
generated. In the simulations, we set the threshold in Definition
\ref{def:edgeset} as $\theta=10$dB and the parameter for RZF as
$\nu=10^{-2}$. We compare the performance of the proposed algorithm
with the following two baselines.

\textbf{Baseline 1 (FFR)}: Fractional frequency reuse (FFR) \cite{Lei_PIMRC07_FFR}
is applied to suppress the inter-cell interference. In each cell,
ZF beamforming is used to serve the users on each subband. 

\textbf{Baseline 2 (Clustered CoMP)}: 3 neighbor BSs form a cluster
and employ cooperative ZF \cite{somekh2009cooperative} to simultaneously
serve all the users within the cluster.

\subsection{Convergence of the Modified Algorithm E}

\begin{figure}
\begin{centering}
\includegraphics[width=85mm]{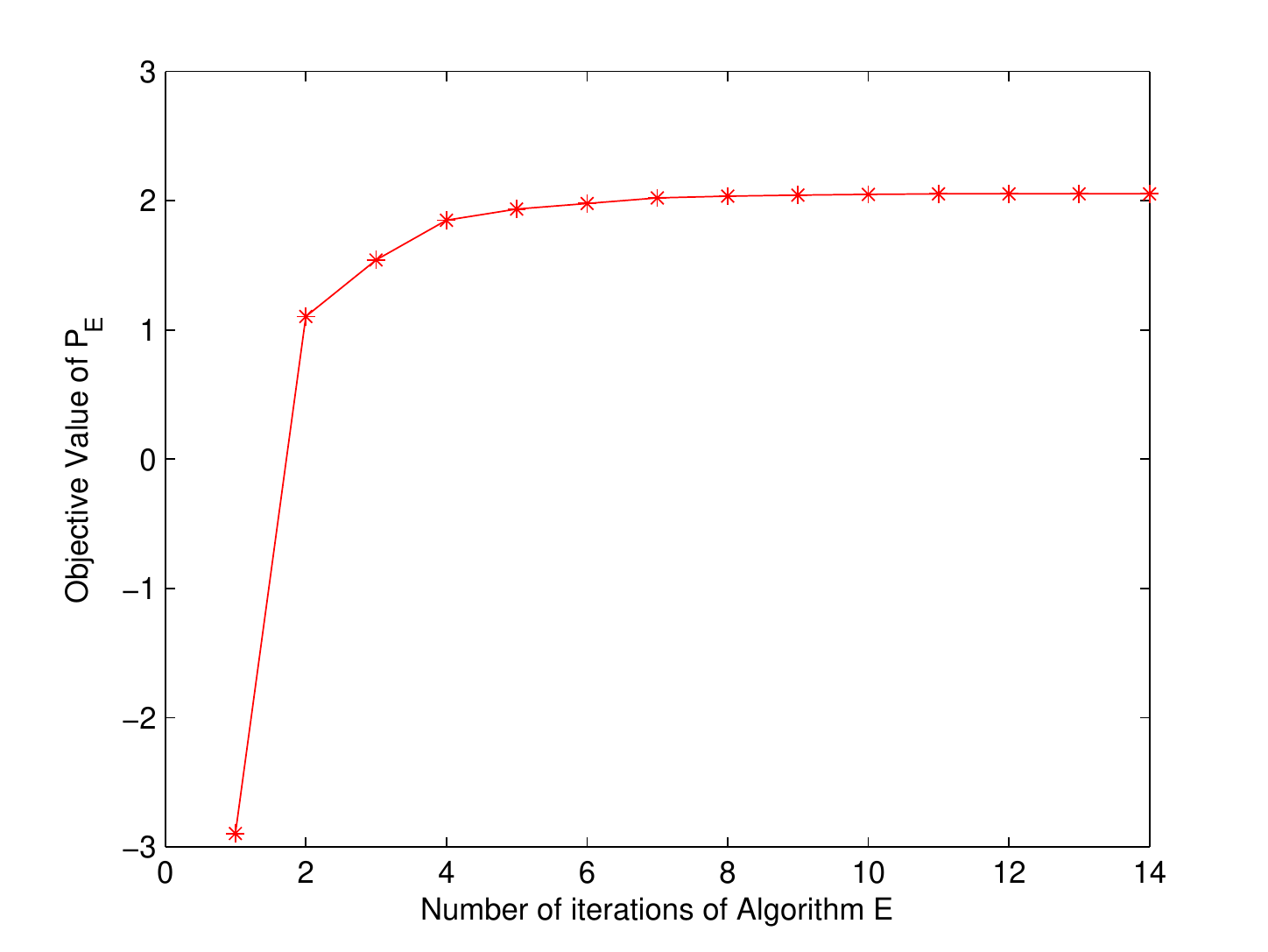}
\par\end{centering}

\protect\caption{\label{fig:convA}{\small{}Objective value of $\mathcal{P}_{E}$ versus
the number of iterations.}}
\end{figure}

Consider the PFS utility\textbf{ }$U\left(\overline{\mathbf{r}}\right)=\frac{1}{K}\sum_{k=1}^{K}\textrm{log}\left(\overline{r}_{k}+\epsilon\right)$\textbf{
}with $\epsilon=10^{-4}$. The per BS transmit power is $P_{c}=10$dB.
In Fig. \ref{fig:convA}, we plot the objective value $U_{E}\left(\Omega\right)$
of $\mathcal{P}_{E}$ versus the number of iterations of the modified
Algorithm E. It can be seen that the modified Algorithm $E$ quickly
converges.

\subsection{Performance Evaluation under PFS Utility}

\begin{figure}
\begin{centering}
\includegraphics[width=85mm]{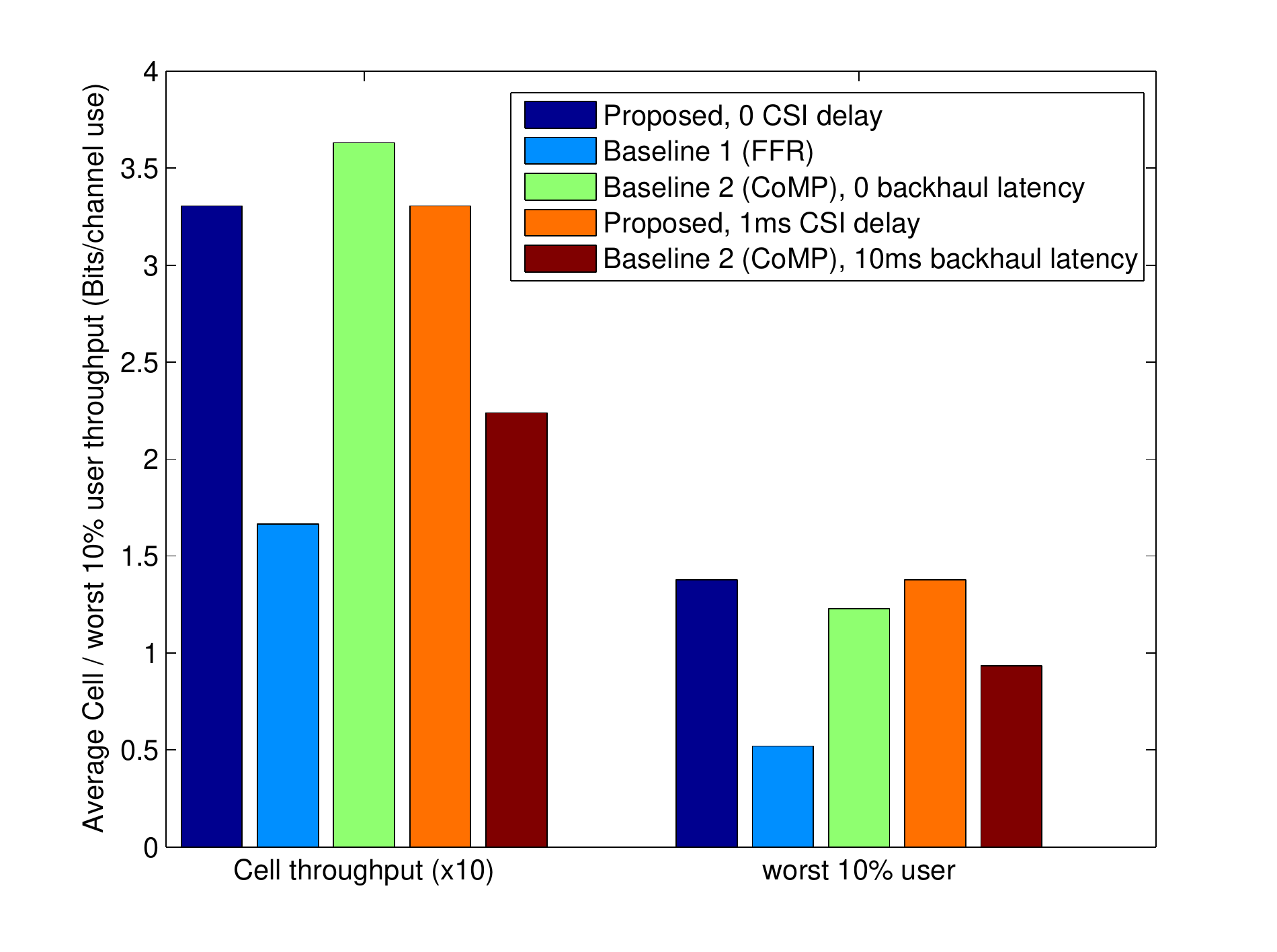}
\par\end{centering}

\protect\caption{\label{fig:bar_throughput}{\small{}Throughput comparisons over different
schemes. The user speed is 3 km/h.}}
\end{figure}

The simulation setup is the same as that in Fig. \ref{fig:convA}.
In Fig. \ref{fig:bar_throughput}, we compare the average cell throughput
of different schemes under different backhaul latencies. For baseline
2, the $3$ cooperative BSs need to exchange CSI and payload data,
and thus there is CSI delay when the backhaul latency is not zero.
When there is CSI delay, the outdated CSI is related to the actual
CSI by the autoregressive model in \cite{Baddour_TWC05_CSIdelaymodel}.
It can be seen that the cell throughput of the proposed scheme is
close to the baseline 2 with zero backhaul latency and is much larger
than baseline 1. The worst 10\% users also benefit from huge throughput
gain over baseline 1. Although the performance of baseline 2 is promising
at zero backhaul latency, the performance quickly degrades at 10ms
backhaul latency. These results demonstrated the superior performance
and the robustness of the proposed hierarchical interference mitigation
w.r.t. signaling latency in backhaul. Table \ref{tab:Cputime} compares
the computational complexity (CPU time) and signaling overhead of
different schemes. The computational complexity and the backhaul signaling
overhead of the proposed scheme are similar to FFR, and are much lower
than CoMP. The real-time CSI estimation overhead of the proposed scheme
is lower than both FFR and CoMP.

\begin{table}
\begin{centering}
{\footnotesize{}}%
\begin{tabular}{|l|l|l|l|}
\hline 
 & {\footnotesize{}CPU } & {\footnotesize{}Backhaul } & {\footnotesize{}Real-time CSI }\tabularnewline
 & {\footnotesize{}time} & {\footnotesize{}signaling } & {\footnotesize{}estimation overhead }\tabularnewline
 &  & {\footnotesize{}overhead} & {\footnotesize{}(Pilot \& CSI feedback)}\tabularnewline
\hline 
{\footnotesize{}Proposed} & {\footnotesize{}0.0260 s} & {\footnotesize{}34.23Mbps} & {\footnotesize{}$\approx$22 PS, 9 $\mathbb{C}^{22}$}\tabularnewline
{\footnotesize{}FFR} & {\footnotesize{}0.0126 s} & {\footnotesize{}16.95Mbps} & {\footnotesize{}48 PS, 12 $\mathbb{C}^{48}$}\tabularnewline
{\footnotesize{}CoMP} & {\footnotesize{}0.1006 s} & {\footnotesize{}111.2Mbps} & {\footnotesize{}48 PS, 12 $\mathbb{C}^{144}$}\tabularnewline
\hline 
\end{tabular}
\par\end{centering}{\footnotesize \par}

{\small{}\protect\caption{\label{tab:Cputime}{\small{}Comparison of the per time slot MATLAB
computational time and per time slot per cell signaling overhead of
different schemes. Assume that the system bandwidth is 1MHz, and the
spatial channel correlation matrices $\mathbf{\Theta}$ changes every
1000 time slots. The other simulation setup is the same as Fig. \ref{fig:bar_throughput}.
The real-time CSI estimation overhead includes the pilot symbol overhead
(in terms of the average number of independent pilot symbols) and
the uplink CSI feedback overhead (in terms of the average number of
feedback channel vectors with different dimensions). For example,
the real-time CSI estimation overhead of the proposed scheme is about
``22 PS, 9 $\mathbb{C}^{22}$'', which means that in average, the
proposed scheme requires transmitting 22 independent pilot symbols
and feedbacking 9 complex channel vectors with average dimension $22$
per time slot per cell.}}
}
\end{table}

\subsection{Performance Evaluation under Sum-rate Utility}

\begin{figure}
\begin{centering}
\includegraphics[width=85mm]{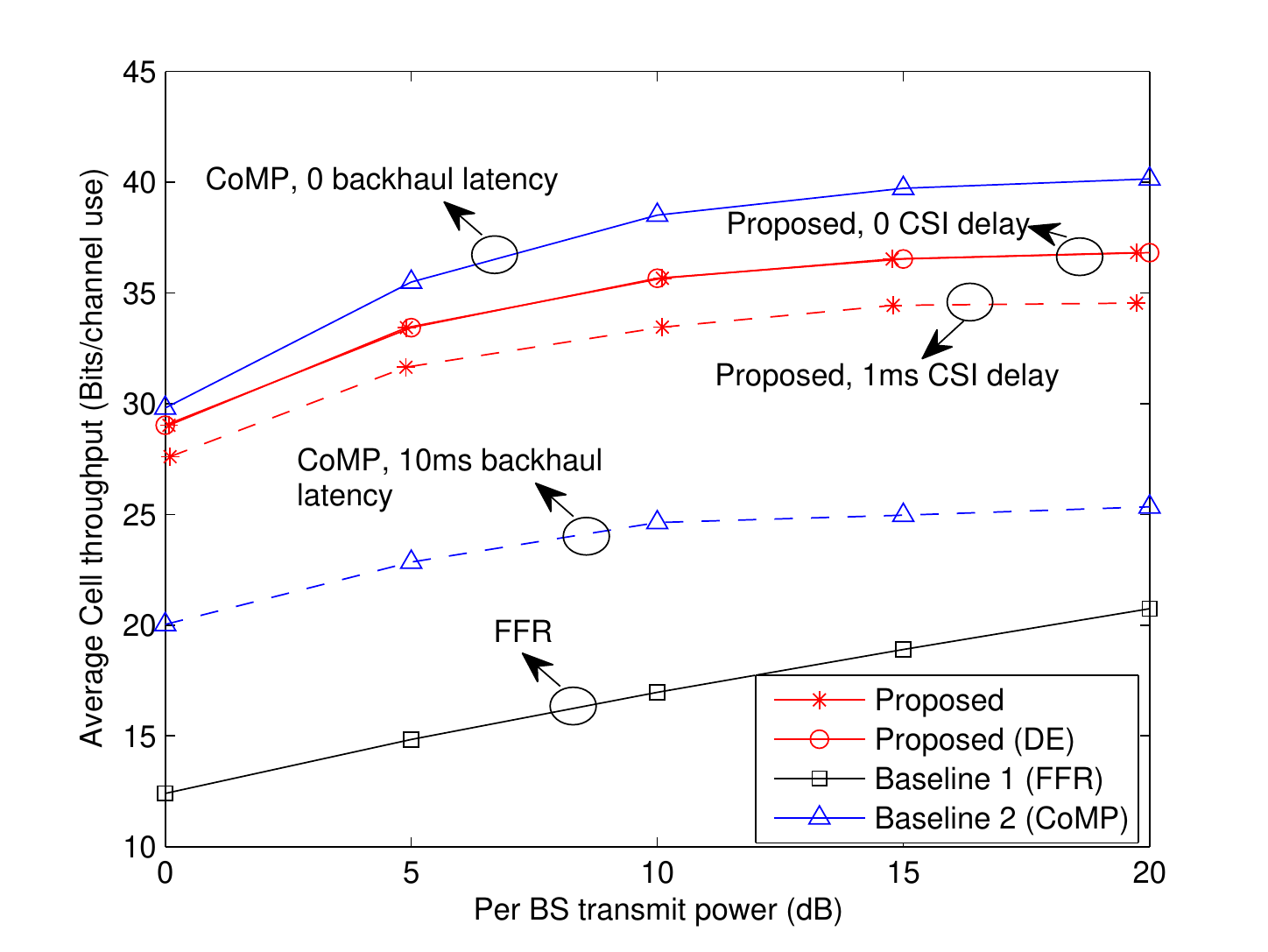}
\par\end{centering}

\protect\caption{\label{fig:sum_throughput}{\small{}Average cell throughput versus
the per BS transmit power $P_{c}$. The user speed is 3 km/h.}}
\end{figure}

Consider the sum-rate utility. In Fig. \ref{fig:sum_throughput},
we plot the average cell throughput $\frac{K}{N}U\left(\overline{\mathbf{r}}\right)$
of different schemes versus the per BS transmit power $P_{c}$. It
can be seen that the cell throughput of the proposed scheme is close
to the baseline 2 with zero backhaul latency and is much larger than
baseline 1. When there is a backhaul latency of 10ms, the proposed
scheme also has a significant throughput gain over baseline 2. The
DE of the cell throughput $\frac{K}{N}U_{E}\left(\Omega\right)$ is
also plotted for the proposed scheme. It can be seen that the DE is
very accurate.

\section{Conclusion\label{sec:Conlusion}}

We propose a hierarchical interference mitigation scheme for massive
MIMO cellular networks. The MIMO precoder is partitioned into \textit{inner
precoder} (for intra-cell interference control) and \textit{outer
precoder} (for inter-cell interference control). We study joint optimization
of the outer precoders, the user selection, and the power allocation.
The optimization only requires the knowledge of spatial correlation
matrices and thus is robust to backhaul latency. We first apply the
random matrix theory to obtain an approximated problem which is non-convex.
Then using the hidden convexity of the problem, we propose \textit{Algorithm
E} to obtain the global optimal solution and a low complexity version
of Algorithm E to find a sub-optimal solution. Simulations show that
the proposed design achieves significant performance gain over various
state-of-the-art baselines.

\appendix

\subsection{Proof of Lemma \ref{lem:AsyWS}\label{sub:Proof-of-LemmaAsyWS}}

Under the zero inter-cell interference constraint in (\ref{eq:ICIZF}),
the $n$-th cell can be viewed as a single-cell downlink system with
equivalent channels $\mathbf{F}_{n}\mathbf{F}_{n}^{\dagger}\mathbf{h}_{k,n},\forall k\in\mathcal{S}_{n}$.
Throughout this proof, the notation $M\rightarrow\infty$ refers to
$M\rightarrow\infty$ such that $0<\underset{M\rightarrow\infty}{\liminf}\left|\mathcal{S}_{n}\right|/M\leq\underset{M\rightarrow\infty}{\limsup}\left|\mathcal{S}_{n}\right|/M<\infty$.
Following similar analysis as in the proof of \cite[Theorem 2]{Wagner_TIT12s_LargeMIMO},
the following lemma can be proved.
\begin{lem}
\label{lem:AsySInr}Let Assumption \ref{asm:LSFM} holds true. As
$M\rightarrow\infty$, we have $r_{k}\left(\Gamma\right)-\hat{r}_{k}^{\circ}\left(\Gamma\right)\overset{a.s}{\rightarrow}0$
and $P_{n}\left(\Gamma\right)-\hat{P}_{n}^{\circ}\left(\Gamma\right)\overset{a.s}{\rightarrow}0$,
where 
\begin{eqnarray}
\hat{r}_{k}^{\circ}\left(\Gamma\right) & = & \textrm{log}\left(1+\frac{p_{k}\xi_{k}^{2}}{\nu^{2}\Upsilon_{k}+\left(\nu+\xi_{k}\right)^{2}}\right),\label{eq:Asyrate1}\\
\hat{P}_{n}^{\circ}\left(\Gamma\right) & = & \frac{1}{M}\sum_{i\in\mathcal{S}_{n}}\frac{p_{i}\nu^{2}e_{i}}{\left(\nu+\xi_{i}\right)^{2}},\label{eq:AsyPow}
\end{eqnarray}
where $\Upsilon_{k}=\frac{1}{M}\sum_{i\in\mathcal{S}_{n}\backslash\left\{ k\right\} }\frac{\nu^{2}p_{i}e_{ik}}{\left(\nu+\xi_{i}\right)^{2}}$;
$\mathbf{e}=\left[e_{i}\right]_{i\in\mathcal{S}_{n}}\in\mathbb{R}^{\left|\mathcal{S}_{n}\right|}$
and $\mathbf{e}_{k}=\left[e_{ki}\right]_{i\in\mathcal{S}_{n}}\in\mathbb{R}^{\left|\mathcal{S}_{n}\right|}$
are given by
\begin{eqnarray}
\mathbf{e} & = & \left(\mathbf{I}-\mathbf{J}\right)^{-1}\mathbf{u},\label{eq:ThetaLE}\\
\mathbf{e}_{k} & = & \left(\mathbf{I}-\mathbf{J}\right)^{-1}\mathbf{u}_{k},\label{eq:ThetaLE1}
\end{eqnarray}
with $\mathbf{J}=\left[J_{ij}\right]_{i\in\mathcal{S}_{n},j\in\mathcal{S}_{n}}\in\mathbb{R}^{\left|\mathcal{S}_{n}\right|\times\left|\mathcal{S}_{n}\right|}$,
$\mathbf{u}=\left[u_{i}\right]_{i\in\mathcal{S}_{n}}\in\mathbb{R}^{\left|\mathcal{S}_{n}\right|}$,
and $\mathbf{u}_{k}=\left[u_{ki}\right]_{i\in\mathcal{S}_{n}}\in\mathbb{R}^{\left|\mathcal{S}_{n}\right|}$
given by
\begin{eqnarray*}
J_{ij}=\frac{\frac{1}{M}\textrm{tr}\tilde{\mathbf{\Theta}}_{i,n}\mathbf{T}_{n}\tilde{\mathbf{\Theta}}_{j,n}\mathbf{T}_{n}}{M\left(\nu+\xi_{j}\right)^{2}},\\
u_{ki}=\frac{1}{\nu^{2}M}\textrm{tr}\tilde{\mathbf{\Theta}}_{i,n}\mathbf{T}_{n}\tilde{\mathbf{\Theta}}_{k,n}\mathbf{T}_{n}, & u_{i}=\frac{1}{\nu^{2}M}\textrm{tr}\tilde{\mathbf{\Theta}}_{i,n}\mathbf{T}_{n}^{2}.
\end{eqnarray*}

\end{lem}

Following similar analysis as in the proof of \cite[Theorem 3]{Wagner_TIT12s_LargeMIMO},
it can be shown that $\Upsilon_{k}=O\left(1\right)$ and $\nu^{2}e_{i}=\xi_{i}+O\left(\nu\right)$.
Then it follows that $\frac{p_{k}\xi_{k}^{2}}{\nu^{2}\Upsilon_{k}+\left(\nu+\xi_{k}\right)^{2}}=p_{k}+O\left(\nu\right)$
and $\hat{P}_{n}^{\circ}\left(\Gamma\right)=\frac{1}{M}\sum_{i\in\mathcal{S}_{n}}\frac{p_{i}}{\xi_{i}}+O\left(\nu\right)$.
From this and Lemma \ref{lem:AsySInr}, Lemma \ref{lem:AsyWS} follows
immediately.

\subsection{Proof of Theorem \ref{thm:AsyEqP}\label{sub:Proof-of-TheoremAsyEqP}}

Let $\Omega^{*}$ be the optimal solution of Problem $\mathcal{P}\left(\mathcal{G}_{T}\right)$.
It can be proved by contradiction that the control policies $\Omega^{*}$
and $\Omega^{\star}$ must satisfy: $\forall j$, $0<\underset{M\rightarrow\infty}{\liminf}\left|\mathcal{S}^{*}\left(j\right)\right|/M\leq\underset{M\rightarrow\infty}{\limsup}\left|\mathcal{S}^{*}\left(j\right)\right|/M<\infty$
and $0<\underset{M\rightarrow\infty}{\liminf}\left|\mathcal{S}^{\star}\left(j\right)\right|/M\leq\underset{M\rightarrow\infty}{\limsup}\left|\mathcal{S}^{\star}\left(j\right)\right|/M<\infty$.
Define two sets{\small{}
\begin{eqnarray*}
\mathcal{B}_{\infty}^{*}\left(j\right) & = & \left\{ n:\:0<\underset{M\rightarrow\infty}{\liminf}\frac{\left|\mathcal{S}_{n}^{*}\left(j\right)\right|}{M}\leq\underset{M\rightarrow\infty}{\limsup}\frac{\left|\mathcal{S}_{n}^{*}\left(j\right)\right|}{M}<\infty\right\} ,\\
\mathcal{B}_{\infty}^{\star}\left(j\right) & = & \left\{ n:\:0<\underset{M\rightarrow\infty}{\liminf}\frac{\left|\mathcal{S}_{n}^{\star}\left(j\right)\right|}{M}\leq\underset{M\rightarrow\infty}{\limsup}\frac{\left|\mathcal{S}_{n}^{\star}\left(j\right)\right|}{M}<\infty\right\} .
\end{eqnarray*}
}Let $\hat{\Omega}^{*}=\left\{ \hat{\Xi}^{*},\mathbf{\hat{q}}^{*}\right\} $
denote a control policy that satisfies $\left|\hat{\Xi}^{*}\right|=\left|\Xi^{*}\right|$,
$\hat{\mathbf{F}}_{n}^{*}\left(j\right)=\mathbf{F}_{n}\left(j\right),\hat{\mathcal{S}}_{n}^{*}\left(j\right)=\mathcal{S}_{n}^{*}\left(j\right),\hat{\mathbf{p}}_{n}^{*}\left(j\right)=\mathbf{p}_{n}^{*}\left(j\right),\forall n\in\mathcal{B}_{\infty}^{*}\left(j\right)$,
and $\hat{\mathcal{S}}_{n}^{*}\left(j\right)=\emptyset,\forall n\notin\mathcal{B}_{\infty}^{*}\left(j\right)$.
Let $\hat{\Omega}^{\star}=\left\{ \hat{\Xi}^{\star},\mathbf{\hat{q}}^{\star}\right\} $
denote a control policy that satisfies $\left|\hat{\Xi}^{\star}\right|=\left|\Xi^{\star}\right|$,
$\hat{\mathbf{F}}_{n}^{\star}\left(j\right)=\mathbf{F}_{n}\left(j\right),\hat{\mathcal{S}}_{n}^{\star}\left(j\right)=\mathcal{S}_{n}^{\star}\left(j\right),\hat{\mathbf{p}}_{n}^{\star}\left(j\right)=\mathbf{p}_{n}^{\star}\left(j\right),\forall n\in\mathcal{B}_{\infty}^{\star}\left(j\right)$,
and $\hat{\mathcal{S}}_{n}^{\star}\left(j\right)=\emptyset,\forall n\notin\mathcal{B}_{\infty}^{\star}\left(j\right)$.
It can be shown that as $M\rightarrow\infty$, we have
\begin{equation}
U\left(\overline{\mathbf{r}}\left(\hat{\Omega}|\mathbf{\Theta}\right)\right)\rightarrow U\left(\overline{\mathbf{r}}\left(\Omega|\mathbf{\Theta}\right)\right),\: U_{E}\left(\hat{\Omega}\right)\rightarrow U_{E}\left(\Omega\right).\label{eq:Omgconv}
\end{equation}
for $\Omega=\Omega^{*},\hat{\Omega}=\hat{\Omega}^{*}$ or $\Omega=\Omega^{\star},\hat{\Omega}=\hat{\Omega}^{\star}$.

For composite control variable $\Gamma$ satisfying the conditions
in Lemma \ref{lem:AsyWS}, it can be shown that $r_{k}\left(\Gamma\right)$
and $P_{n}\left(\Gamma\right)$ are uniformly integrable \cite{David_CUP97_Prob}
w.r.t. $M$. Together with Lemma \ref{lem:AsyWS}, it follows that
\begin{eqnarray}
\underset{M\rightarrow\infty}{\textrm{lim }}\left|\textrm{E}\left[\left.r_{k}\left(\Gamma\right)\right|\mathbf{\Theta}\right]-r_{k}^{\circ}\left(\Gamma|\mathbf{\Theta}\right)\right| & \leq & O\left(\nu\right),\label{eq:Rgap0A}\\
\underset{M\rightarrow\infty}{\textrm{lim }}\left|\textrm{E}\left[\left.P_{n}\left(\Gamma\right)\right|\mathbf{\Theta}\right]-P_{n}^{\circ}\left(\Gamma|\mathbf{\Theta}\right)\right| & \leq & O\left(\nu\right),\label{eq:Pgap0A}
\end{eqnarray}
By definition, we have
\begin{eqnarray}
P_{n}^{\circ}\left(\hat{\Gamma}_{j}^{\star}|\mathbf{\Theta}\right)-P_{c} & \leq & 0,\forall j\label{eq:Pgap0}
\end{eqnarray}
Then it follows from (\ref{eq:Pgap0A}) and (\ref{eq:Pgap0}) that
\begin{equation}
\textrm{E}\left[\left.P_{n}\left(\hat{\Gamma}_{j}^{\star}\right)\right|\mathbf{\Theta}\right]-P_{c}\leq O\left(\nu\right),\:\textrm{as}\: M\rightarrow\infty.\label{eq:Pgapf1}
\end{equation}
Similarly, it can be shown that 
\begin{equation}
P_{n}^{\circ}\left(\hat{\Gamma}_{j}^{*}|\mathbf{\Theta}\right)-P_{c}\leq O\left(\nu\right),\:\textrm{as}\: M\rightarrow\infty.\label{eq:Pgapf2}
\end{equation}
We expand $U\left(\overline{\mathbf{r}}\left(\hat{\Omega}^{*}|\mathbf{\Theta}\right)\right)-U\left(\overline{\mathbf{r}}\left(\hat{\Omega}^{\star}|\mathbf{\Theta}\right)\right)$
as follows{\footnotesize{}
\begin{eqnarray}
U\left(\overline{\mathbf{r}}\left(\hat{\Omega}^{*}|\mathbf{\Theta}\right)\right)-U\left(\overline{\mathbf{r}}\left(\hat{\Omega}^{\star}|\mathbf{\Theta}\right)\right)=\left[U_{E}\left(\hat{\Omega}^{*}\right)-U_{E}\left(\hat{\Omega}^{\star}\right)\right]+\nonumber \\
\left[U\left(\overline{\mathbf{r}}\left(\hat{\Omega}^{*}|\mathbf{\Theta}\right)\right)-U_{E}\left(\hat{\Omega}^{*}\right)\right]+\left[U_{E}\left(\hat{\Omega}^{\star}\right)-U\left(\overline{\mathbf{r}}\left(\hat{\Omega}^{\star}|\mathbf{\Theta}\right)\right)\right].\label{eq:gapI}
\end{eqnarray}
}From (\ref{eq:Rgap0A}), we have 
\begin{equation}
\left|\overline{r}_{k}\left(\Omega|\mathbf{\Theta}\right)-\sum_{j=1}^{\left|\Xi\right|}q_{j}r_{k}^{\circ}\left(\Gamma_{j}|\mathbf{\Theta}\right)\right|\leq O\left(\nu\right),\:\textrm{as}\: M\rightarrow\infty.\label{eq:romga}
\end{equation}
for $\Omega\in\left\{ \hat{\Omega}^{*},\hat{\Omega}^{\star}\right\} $.
Then it follows from (\ref{eq:romga}) and $w_{k}=O\left(1/K\right),\:\forall k$
that 
\begin{eqnarray}
\left|U\left(\overline{\mathbf{r}}\left(\hat{\Omega}^{*}|\mathbf{\Theta}\right)\right)-U_{E}\left(\hat{\Omega}^{*}\right)\right|\leq O\left(\nu\right), & \textrm{as}\: M\rightarrow\infty,\nonumber \\
\left|U_{E}\left(\hat{\Omega}^{\star}\right)-U\left(\overline{\mathbf{r}}\left(\hat{\Omega}^{\star}|\mathbf{\Theta}\right)\right)\right|\leq O\left(\nu\right), & \textrm{as}\: M\rightarrow\infty.\label{eq:gapI0}
\end{eqnarray}
From (\ref{eq:Omgconv},\ref{eq:Pgapf2}) and the definition of $\Omega^{\star}$
and $\Omega^{*}$, we have
\begin{equation}
U_{E}\left(\hat{\Omega}^{*}\right)-U_{E}\left(\hat{\Omega}^{\star}\right)\leq O\left(\nu\right),\label{eq:Iequ}
\end{equation}
Then it follows from (\ref{eq:Omgconv},\ref{eq:gapI},\ref{eq:gapI0},\ref{eq:Iequ})
that
\[
U^{*}-U\left(\overline{\mathbf{r}}\left(\Omega^{\star}|\mathbf{\Theta}\right)\right)\leq O\left(\nu\right),\:\textrm{as}\: M\rightarrow\infty.
\]
This completes the proof for Theorem \ref{thm:AsyEqP}.

\subsection{Proofs for the Results in Section \ref{sub:Global-Optimality-Condition}\label{sub:Proofs-for-the-GOP}}

\subsubsection*{Proof of Lemma \ref{lem:Convexity-of-R}}

Clearly, $\mathcal{R}\subseteq\textrm{Conv}\left(\mathcal{R}^{\textrm{F}}\right)$.
Hence, we only need to prove that any Pareto boundary point $\mathbf{r}^{\star}$
of $\textrm{Conv}\left(\mathcal{R}^{\textrm{F}}\right)$ must lie
in $\mathcal{R}$. First, it is easy to see that $\mathbf{r}^{\star}$
can always be expressed as a convex combination of $K+1$ points $\left\{ \mathbf{r}^{\circ}\left(\Gamma_{1}\right),...,\mathbf{r}^{\circ}\left(\Gamma_{K+1}\right)\right\} $
in $\mathcal{R}^{\textrm{F}}$, i.e., $\mathbf{r}^{\star}=\sum_{j=1}^{K+1}q_{j}\mathbf{r}^{\circ}\left(\Gamma_{j}\right)$,
where $\sum_{j=1}^{K+1}q_{j}=1,q_{j}\in\left[0,1\right]$ and $\Gamma_{j}\in\Xi^{\textrm{F}\circ}\left(P_{c}\right),\forall j$.
Second, $\mathbf{r}^{\circ}\left(\Gamma_{j}\right),\forall j\in\mathcal{J}^{\star}\triangleq\left\{ j:q_{j}>0\right\} $
must lie in the supporting hyperplane to $\textrm{Conv}\left(\mathcal{R}^{\textrm{F}}\right)$
at the Pareto boundary point $\mathbf{r}^{\star}$. Otherwise, $\mathbf{r}^{\star}$
cannot be a Pareto boundary point of $\textrm{Conv}\left(\mathcal{R}^{\textrm{F}}\right)$.
The above two facts imply that $\mathbf{r}^{\star}$ can be expressed
as a convex combination of $K^{'}\leq K$ points in the set $\left\{ \mathbf{r}^{\circ}\left(\Gamma_{1}\right),...,\mathbf{r}^{\circ}\left(\Gamma_{K+1}\right)\right\} $,
i.e., $\mathbf{r}^{\star}=\sum_{j=1}^{K^{'}}q_{j}^{'}\mathbf{r}^{\circ}\left(\Gamma_{j}^{'}\right)$,
where $\sum_{j=1}^{K^{'}}q_{j}^{'}=1,q_{j}^{'}\in\left[0,1\right]$
and $\Gamma_{j}^{'}\in\left\{ \Gamma_{1},...,\Gamma_{K+1}\right\} $.
Hence, $\mathbf{r}^{\star}$ must lie in $\mathcal{R}$.

\subsubsection*{Proof of Lemma \ref{lem:Hiddencvx}}

The first part of Lemma \ref{lem:Hiddencvx} follows directly from
the definition of problem (\ref{eq:EquPLApre-1}) and $\mathcal{P}_{E}\left(\mathcal{G}_{T}\right)$.
The second part of Lemma \ref{lem:Hiddencvx} can be proved by contradiction.
Suppose $\Omega^{\star}$ satisfies $\overline{\mathbf{r}}^{\circ}\left(\Omega^{\star}\right)=\overline{\mathbf{r}}^{\circ\star}$
but is not the global optimal solution of $\mathcal{P}_{E}\left(\mathcal{G}_{T}\right)$.
Then there exists a control policy $\Omega\in\Lambda^{\circ}\left(P_{c}\right)$
such that $U_{E}\left(\Omega\right)>U_{E}\left(\Omega^{\star}\right)$.
Then compared to $\overline{\mathbf{r}}^{\circ\star}$, $\overline{\mathbf{r}}^{\circ}\left(\Omega\right)\in\mathcal{R}$
achieves a larger objective value for problem (\ref{eq:EquPLApre-1}),
which contradicts with the assumption that $\overline{\mathbf{r}}^{\circ\star}$
is the optimal solution of problem (\ref{eq:EquPLApre-1}).

\subsubsection*{Proof of Theorem \ref{thm:Global-Optimality-ConditionPE}}

Suppose $\Omega^{\star}=\left\{ \Xi^{\star},\mathbf{q}^{\star}\right\} $
with $\Xi^{\star}=\left\{ \Gamma_{1}^{\star},...,\Gamma_{\left|\Xi^{\star}\right|}^{\star}\right\} $
satisfies the optimality condition in Theorem \ref{thm:Global-Optimality-ConditionPE}.
It follows from (\ref{eq:OptcondPE}) that 
\begin{equation}
\sum_{k=1}^{K}\mu_{k}^{\star}\left(r_{k}^{\circ}\left(\Gamma_{1}^{\star}\right)-x_{k}\right)\geq0,\:\forall\mathbf{x}\in\mathcal{R},\label{eq:r1x}
\end{equation}
and $\sum_{k=1}^{K}\mu_{k}^{\star}r_{k}^{\circ}\left(\Gamma_{j}^{\star}\right)=\sum_{k=1}^{K}\mu_{k}^{\star}r_{k}^{\circ}\left(\Gamma_{1}^{\star}\right),\forall j$.
Using the above fact and noting that $\overline{r}_{k}^{\circ}\left(\Omega^{\star}\right)=\sum_{j=1}^{\left|\Xi\right|}q_{j}^{\star}r_{k}^{\circ}\left(\Gamma_{j}^{\star}\right)$,
where $q_{j}^{\star}\in\left[0,1\right],\forall j\:\textrm{and}\:\sum_{j=1}^{\left|\Xi\right|}q_{j}^{\star}=1$,
we have
\begin{eqnarray}
\sum_{k=1}^{K}\mu_{k}^{\star}\overline{r}_{k}^{\circ}\left(\Omega^{\star}\right) & = & \sum_{k=1}^{K}\mu_{k}^{\star}\sum_{j=1}^{\left|\Xi\right|}q_{j}^{\star}r_{k}^{\circ}\left(\Gamma_{j}^{\star}\right)\nonumber \\
 & = & \sum_{j=1}^{\left|\Xi\right|}q_{j}^{\star}\sum_{k=1}^{K}\mu_{k}^{\star}r_{k}^{\circ}\left(\Gamma_{j}^{\star}\right)=\sum_{k=1}^{K}\mu_{k}^{\star}r_{k}^{\circ}\left(\Gamma_{1}^{\star}\right).\nonumber \\
\label{eq:rlst}
\end{eqnarray}
Combining (\ref{eq:r1x}) and (\ref{eq:rlst}), we have\textit{
\begin{equation}
\nabla^{T}U\left(\overline{\mathbf{r}}^{\circ}\left(\Omega^{\star}\right)\right)\left(\overline{\mathbf{r}}^{\circ}\left(\Omega^{\star}\right)-\mathbf{x}\right)\geq0,\forall\mathbf{x}\in\mathcal{R}.\label{eq:OptCondHD-1}
\end{equation}
}By Lemma \ref{lem:First-Order-OptimalityHD}, $\overline{\mathbf{r}}^{\circ}\left(\Omega^{\star}\right)$
is the optimal solution of problem (\ref{eq:EquPLApre-1}). Then it
follows from Lemma \ref{lem:Hiddencvx} that $\Omega^{\star}$ is
the global optimal solution of $\mathcal{P}_{E}\left(\mathcal{G}_{T}\right)$.

On the other hand, suppose $\Omega^{\star}$ is the optimal solution
of $\mathcal{P}_{E}\left(\mathcal{G}_{T}\right)$. By Lemma \ref{lem:Hiddencvx},
$\overline{\mathbf{r}}^{\circ}\left(\Omega^{\star}\right)$ is the
optimal solution of (\ref{eq:EquPLApre-1}). Then by Lemma \ref{lem:First-Order-OptimalityHD},
$\Omega^{\star}$ satisfies (\ref{eq:OptCondHD-1}), from which it
can be shown that $\Omega^{\star}$ satisfies the optimality condition
in (\ref{eq:OptcondPE}).

\subsection{Proof of Theorem \ref{thm:Char-of-PW}\label{sub:Proof-of-TheoremCharPW}}

It can be seen that the optimal solution of the following WSRM problem
satisfies (\ref{eq:OptcondPE-1}) 
\begin{equation}
\mathcal{P}_{W}\left(\mathcal{G}_{T},\boldsymbol{\mu}\right):\:\underset{\Gamma}{\textrm{max}}\:\sum_{k=1}^{K}\mu_{k}r_{k}^{\circ}\left(\Gamma\right),\:\textrm{s.t.}\:\Gamma\in\Xi^{\textrm{F}\circ}\left(P_{c}\right).\label{eq:PW}
\end{equation}
Hence, we only need to prove that the output $\Gamma^{\star}\left(\boldsymbol{\mu}\right)$
of Procedure $\textrm{W}^{\star}$ is the optimal solution of $\mathcal{P}_{W}\left(\mathcal{G}_{T},\boldsymbol{\mu}\right)$.

First, we show that $\mathcal{P}_{W}\left(\mathcal{G}_{T},\boldsymbol{\mu}\right)$
is equivalent to a joint user selection and power allocation problem.
\begin{lem}
[Equivalence of $\mathcal{P}_{W}\left(\mathcal{G}_{T},\boldsymbol{\mu}\right)$]\label{lem:Equivalence-of-PW}Let
$\mathcal{S}^{\star},\mathbf{p}^{\star}$ denote an optimal solution
of
\begin{equation}
\underset{\mathcal{S},\mathbf{p}}{\textrm{max}}\sum_{k\in\mathcal{S}}\mu_{k}\textrm{log}\left(1+p_{k}\right),\:\textrm{s.t.}\:\frac{1}{M}\sum_{i\in\mathcal{S}_{n}}\frac{p_{i}}{\xi_{i}}\leq P_{c},\forall n.\label{eq:WSRMP}
\end{equation}
Then $\Gamma^{\star}=\left\{ \mathbf{F}^{\star},\mathcal{S}^{\star},\mathbf{p}^{\star}\right\} $
is an optimal solution of $\mathcal{P}_{W}\left(\mathcal{G}_{T},\boldsymbol{\mu}\right)$,
where $\mathbf{F}^{\star}=\left\{ \mathbf{F}_{1}^{\star},...,\mathbf{F}_{N}^{\star}\right\} $
with $\mathbf{F}_{n}^{\star}=\textrm{orth}\left(\left(\mathbf{I}_{M}-\mathbf{U}\left(\overline{\mathcal{S}}_{n}^{\star}\right)\mathbf{U}^{\dagger}\left(\overline{\mathcal{S}}_{n}^{\star}\right)\right)\sum_{k\in\mathcal{S}_{n}^{\star}}\mathbf{\Theta}_{k,n}\right)$;
and $\overline{\mathcal{S}}_{n}^{\star}=\overline{\mathcal{U}}_{n}\cap\mathcal{S}^{\star}$.
\end{lem}

\begin{IEEEproof}
Lemma \ref{lem:Equivalence-of-PW} can be proved by contradiction.
First, it is easy to see that $\Gamma^{\star}=\left\{ \mathbf{F}^{\star},\mathcal{S}^{\star},\mathbf{p}^{\star}\right\} $
is a feasible solution of $\mathcal{P}_{W}\left(\mathcal{G}_{T},\boldsymbol{\mu}\right)$,
i.e., $\Gamma^{\star}\in\Xi^{\textrm{F}\circ}\left(P_{c}\right)$.
Suppose that $\Gamma^{\star}$ is not an optimal solution of $\mathcal{P}_{W}\left(\mathcal{G}_{T},\boldsymbol{\mu}\right)$.
Then there exists $\Gamma=\left\{ \mathbf{F},\mathcal{S},\mathbf{p}\right\} \in\Xi^{\textrm{F}\circ}\left(P_{c}\right)$
such that $\sum_{k=1}^{K}\mu_{k}r_{k}^{\circ}\left(\Gamma\right)>\sum_{k=1}^{K}\mu_{k}r_{k}^{\circ}\left(\Gamma^{\star}\right)$.
Since $\mathbf{F}$ satisfies the zero inter-cell interference constraint
in (\ref{eq:ICIZF}), we must have $\textrm{span}\left(\mathbf{F}_{n}\mathbf{F}_{n}^{\dagger}\right)\subseteq\textrm{span}\left(\mathbf{I}_{M}-\mathbf{U}\left(\overline{\mathcal{S}}_{n}\right)\mathbf{U}^{\dagger}\left(\overline{\mathcal{S}}_{n}\right)\right),\forall n$.
Let $\overline{\Gamma}=\left\{ \overline{\mathbf{F}},\mathcal{S},\mathbf{p}\right\} $,
where $\overline{\mathbf{F}}=\left\{ \overline{\mathbf{F}}_{1},...,\overline{\mathbf{F}}_{N}\right\} $
with $\overline{\mathbf{F}}_{n}=$$\textrm{orth}\left(\left(\mathbf{F}_{n}\mathbf{F}_{n}^{\dagger}\right)\sum_{k\in\mathcal{S}_{n}}\mathbf{\Theta}_{k,n}\right)$.
It can be shown that $r_{k}^{\circ}\left(\overline{\Gamma}\right)=\textrm{log}\left(1+p_{k}\right)=r_{k}^{\circ}\left(\Gamma\right),\forall k$
and $P_{n}^{\circ}\left(\overline{\Gamma}\right)=P_{n}^{\circ}\left(\Gamma\right)$.
Let $\Gamma^{'}=\left\{ \mathbf{F}^{'},\mathcal{S},\mathbf{p}\right\} $,
where $\mathbf{F}^{'}=\left\{ \mathbf{F}_{1}^{'},...,\mathbf{F}_{N}^{'}\right\} $
with $\mathbf{F}_{n}^{'}=\textrm{orth}\left(\left(\mathbf{I}_{M}-\mathbf{U}\left(\overline{\mathcal{S}}_{n}\right)\mathbf{U}^{\dagger}\left(\overline{\mathcal{S}}_{n}\right)\right)\sum_{k\in\mathcal{S}_{n}}\mathbf{\Theta}_{k,n}\right)$.
It is easy to see that $\Gamma^{'}$ satisfies (\ref{eq:ICIZF}) and
$r_{k}^{\circ}\left(\Gamma^{'}\right)=\textrm{log}\left(1+p_{k}\right)=r_{k}^{\circ}\left(\Gamma\right),\forall k$.
Using the fact that $\textrm{span}\left(\mathbf{F}_{n}\mathbf{F}_{n}^{\dagger}\right)\subseteq\textrm{span}\left(\mathbf{I}_{M}-\mathbf{U}\left(\overline{\mathcal{S}}_{n}\right)\mathbf{U}^{\dagger}\left(\overline{\mathcal{S}}_{n}\right)\right)$,
it can be shown that $P_{n}^{\circ}\left(\Gamma^{'}\right)\leq P_{n}^{\circ}\left(\overline{\Gamma}\right)\leq P_{c}$,
which implies that $\mathcal{S},\mathbf{p}$ is a feasible solution
of Problem (\ref{eq:WSRMP}). Hence, we have $\sum_{k=1}^{K}\mu_{k}r_{k}^{\circ}\left(\Gamma^{\star}\right)\geq\sum_{k=1}^{K}\mu_{k}r_{k}^{\circ}\left(\Gamma^{'}\right)=\sum_{k=1}^{K}\mu_{k}r_{k}^{\circ}\left(\Gamma\right)$,
which contradicts with $\sum_{k=1}^{K}\mu_{k}r_{k}^{\circ}\left(\Gamma\right)>\sum_{k=1}^{K}\mu_{k}r_{k}^{\circ}\left(\Gamma^{\star}\right)$.
This completes the proof.
\end{IEEEproof}

It can be verified that $\mathcal{S}^{\star},\mathbf{p}^{\star}$
in Line 6 of Procedure $\textrm{W}^{\star}$ is the optimal solution
of (\ref{eq:WSRMP}). By Lemma \ref{lem:Equivalence-of-PW}, the output
$\Gamma^{\star}\left(\boldsymbol{\mu}\right)$ of Procedure $\textrm{W}^{\star}$
is the optimal solution of $\mathcal{P}_{W}\left(\mathcal{G}_{T},\boldsymbol{\mu}\right)$.

\subsection{Proofs for the Results in Subsection \ref{sub:Convergence-and-PerformanceE}\label{sub:Proofs-for-theOptAE}}

\subsubsection*{Proof of Lemma \ref{lem:Property-of-AlgorithmE}}

Note that $U_{E}\left(\Omega^{(i+1)}\right)$ is equal to the optimal
value of problem (\ref{eq:fixthetaPdet}) with $\Xi=\widetilde{\Xi}^{(i)}\cup\Gamma^{\star}\left(\boldsymbol{\mu}^{(i+1)}\right)$.
If we restrict $q_{j}=\left(1-\eta\right)\widetilde{q}_{j}^{(i)},j=1,...,\left|\widetilde{\Xi}^{(i)}\right|$,
problem (\ref{eq:fixthetaPdet}) reduces to problem (\ref{eq:maxetaU}).
Hence, $U_{E}\left(\Omega^{(i+1)}\right)$ must be no less than the
optimal value of (\ref{eq:maxetaU}).

\subsubsection*{Proof of Theorem \ref{thm:OptAE}}

Using the fact that any Pareto point of a $K$-dimensional convex
polytope in $\mathbb{R}_{+}^{K}$ can be expressed as a convex combination
of no more than $K$ vertices, it can be shown that there are at most
$K$ non-zero elements in $\mathbf{q}^{(i)}$ in step 1 of Algorithm
E. Hence $\left|\widetilde{\Xi}^{(i)}\right|\leq K,\forall i$ and
the solution found by Algorithm E is feasible. 

For simplicity of notation, let $\overline{\mathbf{r}}^{\circ(i)}=\overline{\mathbf{r}}^{\circ}\left(\Omega^{(i)}\right)$
and $\mathbf{r}^{\circ(i+1)}=\mathbf{r}^{\circ}\left(\Gamma^{\star}\left(\boldsymbol{\mu}^{(i+1)}\right)\right)$.
By Lemma \ref{lem:Property-of-AlgorithmE}, we have $U_{E}\left(\Omega^{(i+1)}\right)\geq U\left(\overline{\mathbf{r}}^{\circ(i)}\right)=U_{E}\left(\Omega^{(i)}\right)$.
Since the objective value is upper bounded, the following lemma holds.
\begin{lem}
\label{lem:Algcp1}Let $\left\{ \Omega^{(i)}\right\} $ be the iterates
generated by Algorithm E. We have $\lim_{i\rightarrow\infty}U_{E}\left(\Omega^{(i)}\right)\rightarrow U^{*}$
for some $U^{*}$. 
\end{lem}

By Assumption \ref{asm:Ufun}, $u\left(r\right)$ is L-Lipschitz,
which implies that $U\left(\mathbf{x}\right)$ is also L-Lipschitz
with the ``L constant'' given by $\widetilde{L}\leq L\max_{k}w_{k}$.
It is well know that the following lemma holds for a L-Lipschitz function.
\begin{lem}
\label{lem:LcOND}If $U\left(\mathbf{x}\right)$ is L-Lipschitz, i.e.,
\[
\left\Vert \nabla U\left(\mathbf{x}\right)-\nabla U\left(\mathbf{y}\right)\right\Vert \leq\widetilde{L}\left\Vert \mathbf{x}-\mathbf{y}\right\Vert ,\forall\mathbf{x}\geq\mathbf{0},\mathbf{y}\geq\mathbf{0},
\]
for some constant $\widetilde{L}>0$, then 
\[
\left|U\left(\mathbf{y}\right)-U\left(\mathbf{x}\right)-\nabla^{T}U\left(\mathbf{x}\right)\left(\mathbf{y}-\mathbf{x}\right)\right|\leq\frac{\widetilde{L}}{2}\left\Vert \mathbf{x}-\mathbf{y}\right\Vert ^{2}.
\]

\end{lem}

Let $\mathbf{d}^{(i+1)}=\mathbf{r}^{\circ(i+1)}-\overline{\mathbf{r}}^{\circ(i)}$
and $\tau_{i}=\nabla^{T}U\left(\overline{\mathbf{r}}^{\circ(i)}\right)\mathbf{d}^{(i+1)}$.
By definition, we have $\tau_{i}\geq0$. With the above two lemmas,
we will show that $\lim_{i\rightarrow\infty}\tau_{i}=0$, which implies
that $U^{*}$ is the global optimal value (this is because $\tau_{i}=0$
means that $\Omega^{(i)}$ satisfies the global optimality condition
in (\ref{eq:OptcondPE})). From Lemma \ref{lem:LcOND}, we have 
\[
U\left(\overline{\mathbf{r}}^{\circ(i)}+\eta\mathbf{d}^{(i+1)}\right)\geq U\left(\overline{\mathbf{r}}^{\circ(i)}\right)+\eta\tau_{i}-\frac{\widetilde{L}\eta^{2}}{2}\left\Vert \mathbf{d}^{(i+1)}\right\Vert ^{2}.
\]
Note that $\left\Vert \mathbf{d}^{(i+1)}\right\Vert ^{2}\leq D$ for
some constant $D>0$ (this is because $\overline{\mathbf{r}}^{\circ(i)},\mathbf{r}^{\circ(i+1)}\in\mathcal{R}$
and $\mathcal{R}$ is clearly a bounded region). Then we have
\begin{equation}
\max_{\eta}U\left(\overline{\mathbf{r}}^{\circ(i)}+\eta\mathbf{d}^{(i+1)}\right)\geq U\left(\overline{\mathbf{r}}^{\circ(i)}\right)+f\left(\tau_{i}\right),\label{eq:U1}
\end{equation}
where $f\left(\tau_{i}\right)\triangleq\max_{\eta\in\left[0,1\right]}\eta\tau_{i}-\frac{\widetilde{L}D\eta^{2}}{2}$
is given by 
\begin{equation}
f\left(\tau_{i}\right)=\begin{cases}
\frac{\tau_{i}^{2}}{2\widetilde{L}D}, & 0\leq\tau_{i}<\widetilde{L}D\\
\tau_{i}-\frac{\widetilde{L}D}{2}, & \tau_{i}\geq\widetilde{L}D.
\end{cases}\label{eq:ftau}
\end{equation}
Clearly, we have
\begin{equation}
\tau_{i}-\frac{\widetilde{L}D}{2}\geq\frac{\tau_{i}}{2},\:\textrm{if}\:\tau_{i}\geq\widetilde{L}D,\label{eq:tau}
\end{equation}
where the equality holds if and only if $\tau_{i}=\widetilde{L}D$.
From (\ref{eq:U1}-\ref{eq:tau}), we have $\max_{\eta}U\left(\overline{\mathbf{r}}^{\circ(i)}+\eta\mathbf{d}^{(i+1)}\right)\geq U\left(\overline{\mathbf{r}}^{\circ(i)}\right)+\min\left(\frac{\tau_{i}}{2},\frac{\tau_{i}^{2}}{2\widetilde{L}D}\right)$.
By Lemma \ref{lem:Property-of-AlgorithmE}, we have $U_{E}\left(\Omega^{(i+1)}\right)\geq U\left(\overline{\mathbf{r}}^{\circ(i)}\right)+\min\left(\frac{\tau_{i}}{2},\frac{\tau_{i}^{2}}{2\widetilde{L}D}\right)=U_{E}\left(\Omega^{(i)}\right)+\min\left(\frac{\tau_{i}}{2},\frac{\tau_{i}^{2}}{2\widetilde{L}D}\right)$.
Hence 
\begin{equation}
U_{E}\left(\Omega^{(i+1)}\right)-U_{E}\left(\Omega^{(i)}\right)\geq\min\left(\frac{\tau_{i}}{2},\frac{\tau_{i}^{2}}{2\widetilde{L}D}\right).\label{eq:UEtao1}
\end{equation}
By Lemma \ref{lem:Algcp1}, we have 
\begin{equation}
\lim_{i\rightarrow\infty}U_{E}\left(\Omega^{(i+1)}\right)-U_{E}\left(\Omega^{(i)}\right)=0.\label{eq:UElim}
\end{equation}
Then it follows from (\ref{eq:UEtao1}) and (\ref{eq:UElim}) that
\begin{eqnarray}
\limsup_{i\rightarrow\infty}\min\left(\frac{\tau_{i}}{2},\frac{\tau_{i}^{2}}{2\widetilde{L}D}\right)\nonumber \\
\leq\limsup_{i\rightarrow\infty}U_{E}\left(\Omega^{(i+1)}\right)-U_{E}\left(\Omega^{(i)}\right)=0.\label{eq:taoi}
\end{eqnarray}
Combining (\ref{eq:taoi}) and the fact that $\tau_{i}\geq0$, we
have $\lim_{i\rightarrow\infty}\tau_{i}=0$. This completes the proof.

\subsection{Proof of Theorem \ref{thm:conve_greedyAE}\label{sub:Proof-of-TheoremgreedyAE}}

Using similar analysis as in the proof of Lemma \ref{lem:Property-of-AlgorithmE},
it can be shown that $U_{E}\left(\Omega^{(i+1)}\right)\geq U_{E}\left(\Omega^{(i)}\right)$
under the modified Algorithm E. Since the objective value is upper
bounded, we have $\lim_{i\rightarrow\infty}U_{E}\left(\Omega^{(i)}\right)\rightarrow\hat{U}$
for some $\hat{U}$. Following similar analysis as that for (\ref{eq:taoi}),
it can be shown that any accumulation point $\left(\hat{\overline{\mathbf{r}}}_{k}^{\circ},\hat{\boldsymbol{\mu}}\right)$
of the iterates $\left\{ \overline{\mathbf{r}}^{\circ}\left(\Omega^{(i)}\right),\boldsymbol{\mu}^{(i+1)}\right\} $
generated by the modified Algorithm E satisfies 
\begin{equation}
\hat{\boldsymbol{\mu}}^{T}\left(\mathbf{r}_{k}^{\circ}\left(\hat{\Gamma}\left(\hat{\boldsymbol{\mu}}\right)\right)-\hat{\overline{\mathbf{r}}}_{k}^{\circ}\right)\leq0.\label{eq:MEpro}
\end{equation}
Moreover, it follows from $\lim_{i\rightarrow\infty}U_{E}\left(\Omega^{(i)}\right)\rightarrow\hat{U}$
that $U\left(\hat{\overline{\mathbf{r}}}_{k}^{\circ}\right)=\hat{U}$. 

Let $\Omega^{\star}$ denote the optimal solution of $\mathcal{P}_{E}\left(\mathcal{G}_{T}\right)$.
Since $\hat{\boldsymbol{\mu}}$ is the gradient of $U\left(\hat{\overline{\mathbf{r}}}_{k}^{\circ}\right)$
(by definition) and $U\left(\mathbf{x}\right)$ is a concave function,
we have
\begin{eqnarray*}
U_{E}^{\star}-\hat{U} & = & U\left(\overline{\mathbf{r}}_{k}^{\circ}\left(\Omega^{\star}\right)\right)-U\left(\hat{\overline{\mathbf{r}}}_{k}^{\circ}\right)\\
 & \leq & \hat{\boldsymbol{\mu}}^{T}\left(\overline{\mathbf{r}}_{k}^{\circ}\left(\Omega^{\star}\right)-\hat{\overline{\mathbf{r}}}_{k}^{\circ}\right),\\
 & \leq & \hat{\boldsymbol{\mu}}^{T}\left(\mathbf{r}_{k}^{\circ}\left(\Gamma^{\star}\left(\hat{\boldsymbol{\mu}}\right)\right)-\mathbf{r}_{k}^{\circ}\left(\hat{\Gamma}\left(\hat{\boldsymbol{\mu}}\right)\right)\right),
\end{eqnarray*}
where the last inequality follows from $\hat{\boldsymbol{\mu}}^{T}\Gamma^{\star}\left(\hat{\boldsymbol{\mu}}\right)\geq\hat{\boldsymbol{\mu}}^{T}\overline{\mathbf{r}}_{k}^{\circ}\left(\Omega^{\star}\right)$
and (\ref{eq:MEpro}).


\end{document}